\documentclass[12pt,letterpaper]{article}
\usepackage[utf8]{inputenc}

\usepackage{color}
\definecolor{darkblue}{rgb}{0.1,0.1,.7}
\usepackage[colorlinks, linkcolor=darkblue, citecolor=darkblue, urlcolor=darkblue, linktocpage]{hyperref}
\usepackage[]{amsmath}
\usepackage[]{graphicx}
\usepackage[]{latexsym}
\usepackage[utf8]{inputenc}
\usepackage{slashed,graphicx,color,amsmath,amssymb}
\usepackage[mathscr]{eucal}
\usepackage{mathrsfs}
\usepackage[margin=10pt,font=small,labelfont=bf]{caption}
\usepackage{amscd}
\usepackage{bm}
\usepackage{xcolor}
\usepackage{upgreek}
\usepackage[square, comma, sort&compress,numbers]{natbib}
\usepackage[all,cmtip]{xy}
\usepackage[margin=1.7in]{geometry}
\usepackage{cleveref}
\usepackage[color=cyan!30!white,linecolor=red,textsize=footnotesize]{todonotes}
\usepackage{soul}
\usepackage{tikz}
\usepackage{pgflibraryarrows}
\usepackage{pgflibrarysnakes}
\usepackage{pgfplots}

\pgfplotsset{compat=1.8}
\tikzset{elegant/.style={smooth,thick,samples=50,magenta}}
\numberwithin{equation}{section}
 \makeatletter
 \g@addto@macro\bfseries{\boldmath}

\makeatletter
\if@todonotes@disabled

\else

\fi
\makeatother

\usepackage{amsmath}
\begin{document}	
\vspace*{-.3in} \thispagestyle{empty}
\begin{flushright}
\end{flushright}
\vspace{.4in} {\Large
\begin{center}
{\bf $T\overline{T}/J\overline{T}$-deformed WZW models from Chern-Simons AdS$_3$ gravity with mixed boundary conditions}
\end{center}}
\vspace{.2in}
\begin{center}
{ Miao He, Yi-hong Gao}
\\
\vspace{.3in}
\small{
\textit{School of Physical Sciences, University of Chinese Academy of Sciences, \\No.19A Yuquan Road, Beijing 100049, China}\\ \vspace{.1cm}
\textit{CAS Key Laboratory of Theoretical Physics, Institute of Theoretical Physics, \\Chinese Academy of Sciences, Beijing 100190, China}
}\\ 
\vspace{.1cm}
\begingroup\ttfamily\small
hemiao@itp.ac.cn,
gaoyh@itp.ac.cn
\endgroup 


\end{center}
\begin{abstract}
\normalsize{In this work we consider AdS$_3$ gravitational theory with certain mixed boundary conditions at infinity. Using the Chern-Simons formalism of AdS$_3$ gravity, we find that these mixed boundary conditions lead to non-trivial boundary terms, which, in turn, produce exactly the spectrum of the $T\bar{T}/J\bar{T}$-deformed CFTs. We then follow the procedure for constructing asymptotic boundary dynamics of AdS$_3$ to derive the constrained $T\bar{T}$-deformed WZW model from Chern-Simons gravity. The resulting theory turns out to be the $T\bar{T}$-deformed Alekseev-Shatashvili action after disentangling the constraints. Furthermore, by adding a $U(1)$ gauge field associated to the current $J$, we obtain one type of the $J\bar T$-deformed WZW model, and show that its action can also be constructed from the gravity side. These results provide a check on the correspondence between the $T\bar{T}/J\bar{T}$-deformed CFTs and the deformations of boundary conditions of AdS$_3$, the latter of which may be regarded as coordinate transformations.}
\end{abstract}


\newpage

\setcounter{page}{1}
\noindent\rule{\textwidth}{.1pt}\vspace{-1.2cm}
\begingroup
\hypersetup{linkcolor=black}
\tableofcontents
\endgroup
\noindent\rule{\textwidth}{.2pt}

\section{Introduction}
Over the past few years, we have seen a surge of interest in deformed 2D conformal field theories~\cite{Smirnov:2016lqw, Cavaglia:2016oda, McGough:2016lol, Giveon:2017nie, Callebaut:2019omt, Tolley:2019nmm, Chakraborty:2019mdf, Chakraborty:2018vja, Apolo:2018qpq, Apolo:2019yfj}. Such theories are integrable, and in some cases allow a holographic description of 3D gravity. So far two kinds of deformations, namely the $T\bar{T}$ deformation and the $J\bar T$ deformation~\cite{Zamolodchikov:2004ce, Cavaglia:2016oda, Guica:2017lia, Anous:2019osb}, have been worked out in detail. It was proposed that the $T\bar{T}$-deformed CFT corresponds to cutoff AdS$_3$ at a finite radius with the Dirichlet boundary condition~\cite{McGough:2016lol, Giribet:2017imm, Kraus:2018xrn}. There are some nontrivial checks on this proposal: the finite size spectrum turns out to be the same as quasilocal energy of the BTZ black hole at finite radius~\cite{McGough:2016lol}, and the $T\bar{T}$ flow equation coincides with the Hamilton-Jacobi equation governing the radial evolution of the classical gravity action in AdS$_3$~\cite{Shyam:2017znq, Donnelly:2019pie}. Based on this proposal, more holographic aspects of the $T\bar{T}$-deformed CFT have been explored, such as entanglement entropy~\cite{Chen:2018eqk, Jeong:2019ylz, Grieninger:2019zts, Jafari:2019qns} and complexity~\cite{Chen:2019mis}. Similarly, the $J\bar T$ deformation also have a holographic interpretation~\cite{Bzowski:2018pcy, Chakraborty:2018vja, Nakayama:2019mvq}. In addition to the above, the torus partition functions of the deformations were studied~\cite{Aharony:2018ics, Aharony:2018bad, Datta:2018thy, Cardy:2018sdv, He:2020cxp}. More recently, the correlation functions of $T\bar{T}$ and $J\bar{T}$ deformations have been computed~\cite{Cardy:2019qao, Guica:2019vnb, He:2019vzf, He:2019ahx, He:2020udl, Li:2020pwa}. As integrable quantum field theories, the deformed 2D CFTs still have infinitely many symmetries. These symmetries have also been studied from 3D gravity perception~\cite{Guica:2019nzm, He:2019glx, Guica:2020uhm}. 
\par
In the context of AdS$_3$/CFT$_2$, the boundary dynamics of AdS$_3$ gravity with the Brown-Henneaux boundary condition turns out to be a $SL(2,\mathbb R)$ WZW model. This result can be derived through the Chern-Simons form of AdS$_3$ gravity. In fact, the AdS$_3$ gravity can be reformulated as a $SL(2,\mathbb R)\times SL(2,\mathbb R)$ Chern-Simons theory, and the Brown-Henneaux boundary condition requires an extra boundary term. The Chern-Simons action with such a boundary term reduces to the sum of two chiral $SL(2,\mathbb R)$ WZW models. Furthermore, this boundary condition also gives certain constraints on the chiral WZW models, which lead to the reduction of the WZW model to the Liouville theory at the classical level~\cite{Coussaert:1995zp} (for more details see the recent review~\cite{Donnay:2016iyk}). More recently, it has been shown that the Chern-Simons AdS$_3$ gravity at quantum level is equivalent to the Alekseev-Shatashvili quantization of coadjoint orbit $\text{Diff}(S^1)/PSL(2, \mathbb R)$ of the Virasoro group~\cite{Cotler:2018zff}. These considerations may be extended to the case of $T\bar{T}$ and $J\bar{T}$ deformation. There already has been some work on this topic, such as using Chern-Simons formalism~\cite{Llabres:2019jtx, Ouyang:2020rpq} to study holographic aspects of $T\bar T/J\bar T$ deformation, as well as the $T\bar T$-deformed Liouville theory~\cite{Leoni:2020rof}. 
\par
In this paper, we focus mainly on the boundary dynamics of AdS$_3$ associated with the $T\bar{T}/J\bar{T}$ deformations. From the cutoff point of view, however, the boundary condition is defined at finite radius, which has no asymptotic degree of freedom. Nevertheless, it is shown that the Dirichlet boundary conditions at finite radius  correspond to the mixed boundary conditions at infinity~\cite{Klebanov:1999tb, Witten:2001ua}. For the $T\bar{T}/J\bar{T}$ deformation, these mixed boundary conditions were obtained in~\cite{Guica:2019nzm, Bzowski:2018pcy} through the variational principle approach. We shall take a close look at these boundary conditions in the Chern-Simons formalism, and derive the nontrivial boundary term. The energy of this system is obtained from the boundary term. As we shall see, these results  agree precisely with the spectra of the $T\bar{T}/J\bar{T}$-deformed CFTs. Moreover, for the $T\bar{T}$ deformation, the total action allows the reduction to the constrained $T\bar{T}$-deformed WZW model. After disentangling the constraints, we show the boundary dynamics are exactly the $T\bar{T}$-deformed Alekseev-Shatashvili action. We will also derive one type of the constrained $J\bar{T}$-deformed WZW model from the gravity side, in which the $U(1)$ current is introduced by adding an extra Abelian gauge field to the Chern-Simons system. The resulting theory is also the $J\bar{T}$-deformed conformal theory. We show that the asymptotic dynamics of AdS$_3$ gravity with the mixed boundary conditions are actually described by the deformed conformal theories. 
\par 
This paper is organized as follows: In section~\ref{TT mixed boundary condition}, we first review the mixed boundary condition of AdS$_3$ for the $T\bar T$ deformation.  After rewriting this boundary condition in the Chern-Simons form, we obtain a nontrivial boundary term. The energy of the whole system can be read off from this boundary term, which matches the finite size spectrum of the $T\bar T$ deformation. In section~\ref{TT deformed WZW model}, the boundary dynamics of AdS$_3$ with mixed boundary condition turns out to be the constrained $T\bar T$-deformed WZW. We also show the equivalence between the sum of two opposite chiral WZW models and the standard non-chiral WZW model under the $T\bar T$ deformation. $J\bar T$ deformation is considered in section~\ref{JT deformation}. Its spectrum is derived from Chern-Simons form by means of the surface integral. The boundary dynamics is also turned out to be a $J\bar T$-deformed conformal theory. Finally, section~\ref{Conclusion and discussion} contains some conclusions and discussions.  
\section{Mixed boundary condition for the $T\bar {T}$ deformation}
\label{TT mixed boundary condition}
In this section, we will study the mixed boundary condition of Chern-Simons AdS$_3$ gravity for the $T\bar{T}$ deformation. We first give a brief review of the mixed boundary condition. Then we put the mixed boundary condition in the Chern-Simons form.  The nontrivial boundary term for mixed boundary condition is obtained. We will also show this boundary term gives exactly the energy of the system, which is in agreement with the spectrum of $T\bar{T}$-deformed CFT. 
\subsection{Review of the mixed boundary condition}
We start from the definition of $T\bar T$-deformed CFT, whose action is given by the $T\bar T$ flow 
\begin{align}
\label{TT-flow}
\frac{\partial S_{T\bar{T}}}{\partial \mu}=\frac{1}{2}\int d^2x\sqrt{\gamma} T\bar T,\quad T\bar{T}=T^{ij}T_{ij}-T^2,
\end{align}
where the metric $\gamma_{ij}$ and stress tensor $T_{ij}$ are defined in the deformed theory. The deformed metric and stress tensor can be expressed in terms of the original ones through the variational principle approach. The basic procedure is to write the variation of the deformed action in terms of the deformed quantities. Then the $T\bar{T}$ flow~\eqref{TT-flow} implies the flow equations 
\begin{align}
\partial_{\mu}\gamma_{ij}=2\hat{T}_{ij},\quad \partial_{\mu}\hat{T}_{ij}=\gamma^{kl}\hat{T}_{ik}\hat{T}_{lj},\quad  \hat T_{ij}=T_{ij}-\gamma_{ij}T^k_k.
\end{align}
Here we mainly draw attention to the flow equation of $\gamma_{ij}$. The solution of $\gamma_{ij}$ flow equation can be expressed as
\begin{align}
\label{g-flow} 
\gamma_{ij}=\gamma_{ij}^{(0)}+2\mu\hat T_{ij}^{(0)}+\mu^2\hat T_{ik}^{(0)}\hat T_{lj}^{(0)}\gamma^{(0)kl},
\end{align} 
where the superscript $(0)$ denotes the quantities of the original theory. \eqref{g-flow} indicates that the background metric of the deformed theory is corrected by the stress tensor of the original theory. If we consider a CFT in the flat spacetime, the deformed theory may not be in the flat spacetime because the background metric is also deformed. This approach was originally developed by Guica and Monten, see~\cite{Bzowski:2018pcy, Guica:2019nzm} for more details.
\par
From the holographic point of view, $\gamma_{ij}$ is interpreted as the boundary metric of AdS$_3$. Therefore, the deformed metric $\gamma_{ij}$ would imply the bulk boundary condition. In general, the solution of 3D gravity can be written in Fefferman-Graham gauge 
\begin{align} 
\label{F-G}
ds^2=\frac{1}{r^2}dr^2+r^2\left(g^{(0)}_{ij}+\frac{1}{r^2}g^{(2)}_{ij}+\frac{1}{r^4}g^{(4)}_{ij}\right)dx^idx^j,
\end{align}
with the constraint
\begin{align} 
\label{metric_constraint}
g^{(4)}_{ij}=\frac{1}{4}g^{(2)}_{ik}g^{(0)kl}g^{(2)}_{jl}.
\end{align}
According to AdS$_3$/CFT$_2$ dictionary, $g^{(2)}_{ij}$ is proportional to the expectation value of the stress tensor of the boundary CFT~\cite{Balasubramanian:1999re}
\begin{align} 
\label{g=t}
g^{(2)}_{ij}=8\pi G(T_{ij}^{(0)}-g^{(0)}_{ij}T^{(0)k}_k)\equiv 8\pi G\hat T_{ij}^{(0)},
\end{align}
where the cosmological constant is set to $\Lambda=-1/\ell^2=-1$. We will use $g_{ij}$ to denote the leading order for the deformed bulk solution. Now, combining \eqref{metric_constraint}, \eqref{g=t} and \eqref{g-flow}, we arrive at the mixed boundary condition\footnote{Here we have redefined the parameter $\mu\sim{\mu}/{8\pi G}$ so that the relation $\mu={1}/{r_c^2}$ holds; this amounts to the choice of units $8\pi G=1$.}
\begin{align} 
\label{TT boundary condition}
g_{ij}=g^{(0)}_{ij}+\mu g^{(2)}_{ij}+\mu^2g^{(4)}_{ij}.
\end{align} 
Namely, the boundary metric of AdS$_3$ is given by~\eqref{TT boundary condition} at infinity. This metric coincides with the boundary metric (expressed within the parentheses in~\eqref{F-G}) at finite radius $r=r_c$, provided the following relation~\cite{McGough:2016lol} is invoked
\begin{align} 
\mu=\frac{1}{r^2_c}.
\end{align} 
This asymptotic behavior allows us to write the bulk solution in the Fefferman-Graham gauge by replacing $g^{(0)}_{ij}$ with  $g_{ij}$. Note that this mixed boundary condition differs in several respects from the Brown-Henneaux boundary condition~\cite{Brown:1986nw}. Although this boundary condition is defined at infinity, the leading order of the  boundary metric $g_{ij}$ is not a flat one. It also breaks the chiral boundary condition in Chern-Simons form. We therefore need a new boundary term to remove inconsistency in the variational principle approach. Besides, the leading order $g_{ij}$ fluctuates, which would inspire us to study the underlying asymptotic dynamics. 
\par
To keep our discussion explicit we consider the Ba\~nados geometry, which constitutes the most general bulk solution of AdS$_3$ with $g^{(0)}_{ij}=\eta_{ij}$. In holomorphic coordinates $(z=\theta+t,\bar z=\theta-t)$, the Ba\~nados metric can be put in the form~\cite{Banados:1998gg}
\begin{align} 
\label{banados}
ds^2=&\frac{dr^2}{r^2}+r^2dzd\bar z+\mathcal L(z)dz^2+\mathcal{\bar L}(\bar z)d\bar z^2+\frac{1}{r^2} \mathcal L(z)\mathcal {\bar L}(\bar z)dzd\bar z,
\end{align} 
where $\mathcal L(z)$ and $\mathcal {\bar L}(\bar z)$ are arbitrary functions depend on $z$ and $\bar z$, respectively. The mixed boundary condition would fix the boundary metric as
\begin{align} 
\label{mixed boundary condition solution}
g_{ij}dx^idx^j=&dzd\bar z+\mu(\mathcal L(z)dz^2+\mathcal{\bar L}(\bar z)d\bar z^2)+\mu^2 \mathcal L(z)\mathcal {\bar L}(\bar z)dzd\bar z.
\end{align} 
Now, introduce the following new coordinates $x^{\pm}$ such that the leading order of the boundary metric takes the manifestly flat form $ds^2_c = dx^{+}dx^{-}$, 
\begin{align} 
\label{transformation}
dx^{+}=dz+\mu\mathcal{ \bar L}(\bar z)d\bar z,\quad dx^{-}=d\bar z+\mu\mathcal{L}(z)dz.
\end{align} 
The deformed bulk solution is obtainable from~\eqref{banados} by performing the inverse of the coordinate transformation
\begin{align}
\label{inverse_transformation} 
dz=\frac{dx^{+}-\mu \mathcal {\bar L}_{\mu}dx^{-}}{1-\mu^2\mathcal L_{\mu}\mathcal {\bar L}_{\mu} },\quad
d\bar z=\frac{dx^{-}-\mu \mathcal {L}_{\mu}dx^{+}}{1-\mu^2\mathcal L_{\mu}\mathcal {\bar L}_{\mu} },
\end{align}
where we used the notations $\mathcal{L}_{\mu}\equiv\mathcal{L}(z(\mu,x^{+},x^{-}))$ and $\mathcal{\bar L}_{\mu}\equiv\mathcal{\bar L}(\bar z(\mu,x^{+},x^{-}))$. The concrete relation between $\mathcal L(x^{+})$ and $\mathcal{L}_{\mu}(x^{+},x^{-})$ may be found in several ways~\cite{Guica:2019nzm}. One of which is that the coordinate transformation~\eqref{transformation} brings the deformed AdS$_3$ solution to the Ba\~nados geometry. The horizon area or energy density should not change under such a coordinate transformation. So comparing these two metrics yields
\begin{align}
\label{LLbar relation} 
\frac{\mathcal{L}_{\mu}(1-\mu \mathcal {\bar L}_{\mu})^2}{(1-\mu^2\mathcal L_{\mu}\mathcal {\bar L}_{\mu})^2}=\mathcal L(x^{+}),\quad
\frac{\mathcal{\bar L}_{\mu}(1-\mu \mathcal {L}_{\mu})^2}{(1-\mu^2\mathcal L_{\mu}\mathcal {\bar L}_{\mu})^2}=\mathcal{\bar L}(x^{-}).
\end{align}
As a result, we can write the deformed AdS$_3$ solution in terms of parameters $\mathcal{L}_{\mu},\mathcal{\bar{L}}_{\mu}$ through the coordinate transformation. 
\par
Moreover, it turns out that the $T\bar T$-deformed theory can be mapped into the original theory via a field dependent coordinate transformation~\cite{Conti:2018tca, Conti:2019dxg}. In terms of the differential form, the coordinate transformation reads
\begin{align} 
\label{TT transformation}
\left(
  \begin{matrix}
   dz\\
    d\bar z\\
  \end{matrix}
  \right)=\frac{1}{1-4\mu^2T(z)\bar T(\bar z)}\left(
  \begin{matrix}
   1&-2\mu T(z)\\
   -2\mu \bar T(\bar z)&1\\
  \end{matrix}
  \right)^{T}
  \left(
  \begin{matrix}
   dx^{+}\\
   dx^{-}\\
  \end{matrix}
  \right).
\end{align}
According to the holographic dictionary, the parameters of Ba\~nados geometry correspond to the stress tensor of the boundary Liouville theory through $\mathcal L(z)=2T(z),\mathcal {\bar L}(\bar z)=2\bar T(\bar z)$~\cite{Banados:1998gg, Donnay:2016iyk}. In this context,  \eqref{inverse_transformation} is consistent with \eqref{TT transformation}. Therefore, we can use the same coordinate transformation in the bulk to get the deformed AdS$_3$ solution.
\subsection{Chern-Simons formalism and the boundary term}
It is well-known that three dimensional Einstein gravity with a negative cosmological constant can be expressed as $SL(2,\mathbb{R})\times SL(2,\mathbb{R})$ Chern-Simons gauge theory~\cite{Witten:1988hc}, whose action is
\begin{align}
\label{cs gravity action}
S(A,\bar A)=I(A)-I(\bar A),
\end{align} 
where 
\begin{align}
\label{CS action}
I(A)=\frac{\kappa}{4\pi}\int_{M}\text{Tr}\left(A\wedge dA+\frac{2}{3}A\wedge A \wedge A \right),\quad \kappa=\frac{1}{4G}.
\end{align} 
The gauge fields $A,\bar A$ valued in two independent copies of $SL(2,\mathbb{R})$, which are defined as the combination of vielbein and spin connection 
\begin{align}
A^a=\omega^a+e^a,\quad\bar A^a=\omega^a-e^a.
\end{align}
The equations of motion are
\begin{align}
\label{equations of motion}
\text{d}A+A\wedge A=0,\quad \text{d}\bar A+\bar A\wedge\bar A=0.
\end{align}
It turns out that these equations are equivalent to first order gravitational field equations. 
\par 
Let us first take a look at the Ba\~nados geometry~\eqref{banados} in Chern-Simons form. The corresponding gauge fields can be calculated
\begin{align}
\label{A}
A=&\frac{1}{r}L_0dr+\left(rL_{-1}+\frac{\mathcal L(z)}{r}L_1\right)dz,\\
\label{Abar}
\bar A=&-\frac{1}{r}L_0dr+\left(\frac{\mathcal {\bar L}(\bar z)}{r}L_{-1}+rL_1\right)d\bar z.
\end{align}
where $L_0, L_{\pm 1}$ are Lie-algebra generators of $SL(2,\mathbb{R})$; see Appendix~\ref{Conventions} for our convention. These gauge fields also can be obtained by solving~\eqref{equations of motion} with the chiral boundary condition $A_{\bar z}=0,\ \bar A_z=0$~\cite{Banados:1998gg}. A useful trick to factor out the boundary degree of freedom is performing the following gauge transformation
\begin{align}
\label{radial gauge 1}
A=b^{-1}(\text{d}+a)b,\quad 
\bar A=b(\text{d}+\bar a)b^{-1},\quad
b=e^{L_0\ln r}=\left(
  \begin{matrix}
   \sqrt{r}& 0\\
   0&\frac{1}{\sqrt r}\\
  \end{matrix}
  \right).
\end{align}
In this case, the reduced connections have the explicit form
\begin{align}
a=\Big(L_{-1}+\mathcal{L}(z)L_1\Big)dz,\quad \bar a=\Big(\mathcal {\bar L}(\bar z)L_{-1}+L_1\Big)d\bar z,  
\end{align}
which depend on the boundary coordinates $(z,\bar{z})$ only. For later discussion, we would like to use the coordinates $\theta=(z+\bar z)/2,\ t=(z-\bar z)/2$ and impose the periodic condition $\theta\sim\theta+R$. Then the chiral boundary condition becomes $A_{t}=A_{\theta}$ and $\bar A_{t}=-\bar A_{\theta}$. 
Now one can go through a consistent variational principle approach by adding some boundary terms to the action. The total action associated to the chiral boundary condition was found in~\cite{Coussaert:1995zp}, which takes the form 
\begin{align}
\label{total B-H boundary term}
S_{\rm{tot}}(A,\bar A)=I(A)-I(\bar A)-\frac{\kappa}{4\pi}\int_{\partial M}dtd\theta\text{Tr}(A_{\theta}^2+\bar A_{\theta}^2).
\end{align} 
In the Hamiltonian formalism, the supplementary boundary term plays the role of a surface integral, which implies the total energy of this system~\cite{Regge:1974zd}. Inserting~\eqref{A} and~\eqref{Abar} into~\eqref{total B-H boundary term}, the boundary term becomes 
\begin{align}
\label{B-H boundary term}
\mathcal B_0=-\frac{\kappa}{2\pi}\int_{\partial M}dtd\theta\Big(\mathcal L(z)+\mathcal {\bar L}(\bar z)\Big).
\end{align} 
For the BTZ black holes, $\mathcal L(z)=\mathcal L_0,\mathcal{\bar L}(\bar z)=\mathcal {\bar L}_0$, the boundary term~\eqref{B-H boundary term} gives exactly the energy (or mass) of the black hole 
\begin{align}
E=\frac{\kappa R}{2\pi}(\mathcal L_0+\mathcal {\bar L}_0)=M.
\end{align}  
\par
We now turn to the investigation of the mixed boundary condition for the $T\bar{T}$ deformation. As we shall see, this mixed boundary condition can be obtained from the Brown-Henneaux boundary condition through a field dependent coordinate transformation~\eqref{inverse_transformation}. Consequently, the gauge fields corresponding to the mixed boundary condition are given by 
\begin{align}
\label{TT deformed A}
\tilde {A}=&\frac{1}{r}L_0dr+\frac{1}{1-\mu^2\mathcal L_{\mu}\mathcal {\bar L}_{\mu}}\left(rL_{-1}+\frac{1}{r}\mathcal {L}_{\mu}L_1\right)(dx^{+}-\mu \mathcal {\bar L}_{\mu}dx^{-}),\\
\label{TT deformed Abar}
\bar{\tilde{A}}=&-\frac{1}{r}L_0dr+\frac{1}{1-\mu^2\mathcal L_{\mu}\mathcal {\bar L}_{\mu}}\left(\frac{1}{r}\mathcal{ \bar L}_{\mu}L_{-1}+rL_{1}\right)(dx^{-}-\mu \mathcal {L}_{\mu}dx^{+}).
\end{align}
We use tilde symbols to denote the quantities in the deformed theory. One can clearly see that the deformed gauge fields obey
\begin{align}
\mu \mathcal {\bar L}_{\mu}\tilde A_{+}+\tilde A_-=0,\quad 
\bar{\tilde A}_{+}+\mu \mathcal {L}_{\mu}\bar{\tilde A}_-=0,
\end{align}
instead of the chiral boundary condition. That is to say,  the mixed boundary condition breaks the chiral boundary condition. However, the equation of motion still holds, because the deformed bulk solution also satisfies Einstein equation. In the coordinates $\tilde\theta=(x^++x^-)/2,\tilde t=(x^+-x^-)/2$, the gauge fields $\tilde A$ and $\bar{\tilde A}$ have the following relations
\begin{align}
\label{AAbar relation}
\tilde A_{\tilde t}=\frac{1+\mu\mathcal {\bar L}_{\mu}}{1-\mu\mathcal {\bar L}_{\mu}}\tilde A_{\tilde \theta},\quad 
\bar {\tilde A}_{\tilde t}=-\frac{1+\mu\mathcal { L}_{\mu}}{1-\mu\mathcal {L}_{\mu}}\bar {\tilde A}_{\tilde \theta}.
\end{align} 
The $r$ dependence of the deformed gauge fields can also be eliminated through the gauge transformation~\eqref{radial gauge 1}. Thus, we get the reduced connections for deformed theory
\begin{align}
\label{a-tt}
\tilde a_{\tilde\theta}=&\frac{1-\mu \mathcal {\bar L}_{\mu}}{1-\mu^2\mathcal {L}_{\mu}\mathcal {\bar L}_{\mu}}(L_{-1}+\mathcal{L}_{\mu}L_1),\quad \tilde a_{\tilde t}=\frac{1+\mu\mathcal {\bar L}_{\mu}}{1-\mu\mathcal {\bar L}_{\mu}}\tilde a_{\tilde \theta},\\
\label{abar-tt}
\bar {\tilde a}_{\tilde\theta}=&\frac{1-\mu \mathcal {L}_{\mu}}{1-\mu^2\mathcal {L}_{\mu}\mathcal {\bar L}_{\mu}}(\mathcal {\bar L}_{\mu}L_{-1}+L_1),\quad \bar{\tilde a}_{\tilde t}=-\frac{1+\mu\mathcal { L}_{\mu}}{1-\mu\mathcal {L}_{\mu}}\bar{\tilde a}_{\tilde \theta}.
\end{align}
This is the mixed boundary condition in Chern-Simons form. In order to have a well-defined variational principle, we have to add a supplementary boundary term. It turns out that the corrected boundary term is
\begin{align}
\label{TT boundary term}
\mathcal B=&-\frac{\kappa}{4\pi}\int_{\partial M}d\tilde td\tilde\theta \left[\frac{1-\mu^2\mathcal { L}_{\mu}\mathcal {\bar L}_{\mu}}{1-\mu\mathcal {\bar L}_{\mu}}\text{Tr}\left(\tilde a_{\tilde\theta}^2\right)+\frac{1-\mu^2\mathcal { L}_{\mu}\mathcal {\bar L}_{\mu}}{1-\mu\mathcal {L}_{\mu}}\text{Tr}\left(\bar{\tilde a}_{\tilde\theta}^2\right)\right]\nonumber\\
=&-\frac{\kappa}{2\pi}\int_{\partial M}d\tilde td\tilde\theta\frac{\mathcal {L}_{\mu}+\mathcal {\bar L}_{\mu}-2\mu\mathcal {L}_{\mu}\mathcal {\bar L}_{\mu}}{1-\mu^2\mathcal { L}_{\mu}\mathcal {\bar L}_{\mu}},
\end{align}
where we have invoked~\eqref{a-tt} and~\eqref{abar-tt} in the last step. The detailed derivation of this nontrivial boundary term is given in Appendix~\ref{variation TTbar}.
\par
Here we give some comments about this boundary term. This term reduces to the limiting case~\eqref{B-H boundary term} when $\mu\to 0$. Unlike the limiting case where the chiral boundary condition holds, the boundary term \eqref{TT boundary term} in general cannot be separated into a chiral part depending only on $\tilde a$ and an antichiral part depending only on $\bar{\tilde{a}}$. One may see this more clearly by writing $\mathcal{L}_{\mu},\mathcal{\bar L}_{\mu}$ in terms of the reduced connections. As a consequence, the chiral action $I(A)$ and the antichiral action $I(\bar A)$ in Chern-Simons theory are coupled to each other through the boundary interaction term~\eqref{TT boundary term}, as long as $\mu\ne 0$. This is the effect of $T\bar T$ deformation in Chern-Simons gravity.  
\par
The boundary term also gives rise to the total energy of this system. Working in the Hamiltonian formalism, the surface integral reads 
\begin{align}
\label{TT energy 1}
E=&\frac{\kappa}{2\pi}\int_{\partial M}d\tilde\theta\frac{\mathcal {L}_{\mu}+\mathcal {\bar L}_{\mu}-2\mu\mathcal {L}_{\mu}\mathcal {\bar L}_{\mu}}{1-\mu^2\mathcal { L}_{\mu}\mathcal {\bar L}_{\mu}},
\end{align} 
which is consistent with the result derived from the bulk stress tensor~\cite{Guica:2019nzm}. For the BTZ black holes,  we can work out the total energy with the help of~\eqref{LLbar relation}
\begin{align}
\label{TT energy2}
E=&\frac{R}{\mu}\left(1-\sqrt{1-\frac{2\mu}{R}M+\frac{\mu^2}{R^2}J^2}\right),
\end{align}  
where $M=R(\mathcal L_{0}+\mathcal {\bar L}_0),J=R(\mathcal L_{0}-\mathcal {\bar L}_0)$ are the mass and the angular momentum of the black hole, respectively. The total energy of this system is in agreement with the spectrum of the $T\bar{T}$-deformed CFT. $E$ precisely matches the quasi-local energy of the BTZ black hole due to $\mu= 1/r_c^2$. This result is consistent with the cutoff point of view~\cite{McGough:2016lol}. However, the mixed boundary condition considered in this paper is actually an asymptotic boundary condition, which is defined at infinity rather than at the finite radius $r=r_c$. The advantage of this mixed boundary condition is that we can study the boundary dynamics directly in Chern-Simons theory, as we shall discuss in the next section. 
\section{From Chern-Simons theory to $T\bar{T}$-deformed WZW model}
\label{TT deformed WZW model}
In this section, we would like to study the boundary dynamics of AdS$_3$ with the certain mixed boundary condition. We first take a short look at the chiral boundary condition . It is shown that the Chern-Simons action can be reduced to the WZW model~\cite{Moore:1989yh}
\begin{align}
I(A)=&\frac{\kappa}{4\pi}\int_{\partial M}dtd\theta\text{Tr}(a_{\theta}a_{t})+\Gamma[G],\quad \Gamma[G]=\frac{\kappa}{12\pi}\int_{M}\text{Tr}[(G^{-1}\text{d}G)^3],
\end{align} 
where $a=g^{-1}\text{d}g$, $A=G^{-1}\text{d}G$ and $\Gamma[G]$ is the Wess-Zumino term. The gauge fields can be written in this form because one can choose the pure gauge solution of the equation of motion~\eqref{equations of motion}. After adding the boundary term~\eqref{B-H boundary term}, the total action~\eqref{total B-H boundary term} could reduce to a sum of two chiral WZW actions 
\begin{align}
\label{two chiral WZW}
S_{\text{tot}}=&\frac{\kappa}{4\pi}\int_{\partial M}\text{Tr}[a_{\theta}(a_{t}-a_{\theta})]+\Gamma[G]
-\frac{\kappa}{4\pi}\int_{\partial M}\text{Tr}[\bar a_{\theta}(\bar a_{t}+\bar a_{\theta})]-\Gamma[\bar{G}].
\end{align}
where $g$ and $G$ take values in $SL(2,\mathbb{R})$. It turns out that~\eqref{two chiral WZW} produces a non-chiral $SL(2,\mathbb R)$ WZW model, and the latter allows a further reduction to the Liouville theory classically~\cite{Forgacs:1989ac}. At the quantum level, the Chern-Simons gravity is equivalent to the Alekseev-Shatashvili quantization of Virasoro group~\cite{Cotler:2018zff}. In other words, the asymptotic dynamics of AdS$_3$ with the Brown-Henneaux boundary condition can be described by the conformally invariant theory.
\par
The above consideration can be extended to the case where the mixed boundary condition is imposed. As we shall see, the corresponding boundary term~\eqref{TT boundary term} leads to a coupling between two opposite chiral WZW models, and the resulting theory is equivalent to the $T\bar{T}$-deformed non-chiral WZW model. Moreover, the mixed boundary condition also gives constraints on the $T\bar{T}$-deformed WZW models, which would give a further reduction to the $T\bar{T}$-deformed Alekseev-Shatashvili action.  
\subsection{Reduction to a sum of two coupled chiral WZW actions}
Given that an action with the well-defined variational principle, we are ready to reduce the Chern-Simons action to the $T\bar{T}$-deformed WZW model. The main difference with the CFT case is the boundary term. Firstly, we would like to express the boundary term in terms of the gauge fields.
In the following we find it is convenient to define 
\begin{align}
X_{ij}\equiv\text{Tr}(\tilde A_{i}\tilde A_{j})=\text{Tr}(\tilde a_{i}\tilde a_{j}),\quad \bar X_{ij}\equiv\text{Tr}(\bar{\tilde A}_{i}\bar{\tilde A}_{j})=\text{Tr}(\bar{\tilde a}_{i}\bar{\tilde a}_{j}).
\end{align} 
According to~\eqref{a-tt} and~\eqref{abar-tt}, one can write $\mathcal{L}_{\mu}$ in terms of $X_{\tilde\theta\tilde\theta}$ and $\bar X_{\tilde\theta\tilde\theta}$
\begin{align}
\label{L-a relation}
\mathcal L_{\mu}=&\frac{\pm[1+\mu(X_{\tilde\theta\tilde\theta}-\bar X_{\tilde\theta\tilde\theta})]\sqrt{1-2\mu(X_{\tilde\theta\tilde\theta}+\bar X_{\tilde\theta\tilde\theta})+\mu^2(X_{\tilde\theta\tilde\theta}-\bar X_{\tilde\theta\tilde\theta})^2}}{2\mu^2X_{\tilde\theta\tilde\theta}}\nonumber\\
&+\frac{1-2\mu\bar X_{\tilde\theta\tilde\theta}+\mu^2(X_{\tilde\theta\tilde\theta}-\bar X_{\tilde\theta\tilde\theta})^2}{2\mu^2X_{\tilde\theta\tilde\theta}},
\end{align} 
as well as a similar expression for $\mathcal {\bar L}_{\mu}$. It is straightforward to derive the following identity
\begin{align}
&\sqrt{1-2\mu\left(X_{\tilde\theta\tilde\theta}+\bar X_{\tilde\theta\tilde\theta}\right)+\mu^2\left(X_{\tilde\theta\tilde\theta}-\bar X_{\tilde\theta\tilde\theta}\right)^2}=1-\mu\left(\frac{1-\mu^2\mathcal {L}_{\mu}\mathcal {\bar L}_{\mu}}{1-\mu\mathcal {\bar L}_{\mu}}X_{\tilde\theta\tilde\theta}+\frac{1-\mu^2\mathcal { L}_{\mu}\mathcal {\bar L}_{\mu}}{1-\mu\mathcal {L}_{\mu}}\bar X_{\tilde\theta\tilde\theta}\right).
\end{align}
Comparing this with the first line of~\eqref{TT boundary term}, the boundary term $\mathcal B$ can be expressed as 
\begin{align}
\mathcal B
=&\frac{\kappa}{4\pi}\int_{\partial M}d\tilde td\tilde \theta\frac{1}{\mu}\left(\sqrt{1-2\mu\left(X_{\tilde\theta\tilde\theta}+\bar X_{\tilde\theta\tilde\theta}\right)+\mu^2\left(X_{\tilde\theta\tilde\theta}-\bar X_{\tilde\theta\tilde\theta}\right)^2}-1\right).
\end{align}
It follows that the total Chern-Simons action consistent with the mixed boundary condition may reduce to 
\begin{align}
\label{TT deformed chiral WZW action}
S_{\rm{total}}=&\frac{\kappa}{4\pi}\int_{\partial M}d\tilde td\tilde \theta\left(X_{\tilde\theta\tilde t}-\bar X_{\tilde\theta\tilde t}\right)+\Gamma[G]-\Gamma[\bar G]\nonumber\\
+&\frac{\kappa}{4\pi}\int_{\partial M}d\tilde td\tilde \theta\frac{1}{\mu}\left(\sqrt{1-2\mu\left(X_{\tilde\theta\tilde\theta}+\bar X_{\tilde\theta\tilde\theta}\right)+\mu^2\left(X_{\tilde\theta\tilde\theta}-\bar X_{\tilde\theta\tilde\theta}\right)^2}-1\right).
\end{align}
This is exactly the $T\bar{T}$-deformed chiral WZW action, which was derived from the $T\bar{T}$ flow equations~\cite{Ouyang:2020rpq}. Here we derive the $T\bar T$-deformed WZW model based on the Chern-Simons AdS$_3$ gravity with the mixed boundary condition. 
\par
In order to see the effect of $T\bar{T}$ deformation, one may expand~\eqref{TT deformed chiral WZW action} as a Taylor series with respect to $\mu$. The first few terms of this expansion read
\begin{align}
\label{TT deformed chiral WZW action-expansion}
S_{\rm{total}}=&\frac{\kappa}{4\pi}\int_{\partial M}d\tilde td\tilde \theta\left[X_{\tilde\theta\tilde t}-X_{\tilde\theta\tilde\theta}-\bar X_{\tilde\theta\tilde t}-\bar X_{\tilde\theta\tilde\theta}\right]+\Gamma[G]-\Gamma[\bar G]\nonumber\\
+&\frac{\kappa\mu}{8\pi}\int_{\partial M}d\tilde td\tilde \theta\left[X_{\tilde\theta\tilde\theta}-\bar X_{\tilde\theta\tilde\theta}-\left(X_{\tilde\theta\tilde\theta}+\bar X_{\tilde\theta\tilde\theta}\right)^2\right]+O(\mu^2).
\end{align}
The leading order reproduces the sum of two decoupled chiral WZW actions, as presented in~\eqref{two chiral WZW}. The deformation contributes to higher order terms of $\mu$. Clearly, such higher order terms can no longer be written as the sum of a left-moving part and a right-moving part. In other words, the $T\bar{T}$ deformation provides a coupling between two opposite chiral degrees of freedom. 
\subsection{Equivalence to $T\bar T$-deformed non-chiral WZW action}
As is well known, the sum of left and right chiral WZW actions is equivalent to the standard non-chiral WZW action~\cite{Coussaert:1995zp}. It is natural to expect that \eqref{TT deformed chiral WZW action} is equivalent to a $T\bar{T}$-deformed version of the non-chiral WZW model. By using the usual technique in~\cite{Coussaert:1995zp, Donnay:2016iyk}, we will verify this in this subsection. First, we combine the gauge fields $g,\bar g$
\begin{align}
k\equiv g^{-1}\bar g,\quad K\equiv G^{-1}\bar G,
\end{align}
and introduce the new variables
\begin{align}
\Pi=&-\bar g^{-1}\partial_{\tilde\theta}gg^{-1}\bar g-\bar g^{-1}\partial_{\tilde\theta}\bar g,\\
k^{-1}\partial_{\tilde t}k=&-\bar g^{-1}\partial_{\tilde t}gg^{-1}\bar g+\bar g^{-1}\partial_{\tilde t}\bar g,\\
k^{-1}\partial_{\tilde\theta}k=&-\bar g^{-1}\partial_{\tilde\theta}gg^{-1}\bar g+\bar g^{-1}\partial_{\tilde\theta}\bar g.
\end{align} 
The sum of Wess-Zumino terms becomes 
\begin{align}
\Gamma[G]-\Gamma[\bar G]=-\Gamma[K]+\int_{\partial M}\text{Tr}\left(\text{d}\bar g\bar g^{-1}\text{d}gg^{-1}\right).
\end{align}
We then write the $T\bar T$-deformed chiral WZW action~\eqref{TT deformed chiral WZW action} in terms of the new variables $\Pi$ and $k^{-1}\text{d}k$
\begin{align}
\label{action_pi_k}
S[\Pi,k]=\frac{\kappa}{4\pi}\int_{\partial M}\left[\text{Tr}(\Pi\dot k)+\frac{1}{\mu}\left(\sqrt{1-\mu\text{Tr}\left(k'^2+\Pi^2\right)+\mu^2\text{Tr}\left(k'\Pi\right)\text{Tr}\left(k'\Pi\right)}-1\right)\right]-\Gamma[K],
\end{align}
where we used the notation $k'=k^{-1}\partial_{\tilde\theta}k$ and $\dot k=k^{-1}\partial_{\tilde t}k$. 
\par
The auxiliary variable $\Pi$ can be eliminated by the equation of motion. Varying the action~\eqref{action_pi_k} with respect to $\Pi$, we obtain the equation of motion 
\begin{align}
\dot k=\frac{\Pi-\mu\text{Tr}\left(k'\Pi\right)k'}{\sqrt{\Omega}},\quad \Omega=&1-\mu\left[\text{Tr}\left(k'^2\right)+\text{Tr}\left(\Pi^2\right)\right]+\mu^2\left[\text{Tr}(k'\Pi)\right]^2,
\end{align}
where $\Omega$ is introduced for convenience. According to the above equation, we get the relations
\begin{align}
\text{Tr}(\dot k\Pi)=&\frac{\text{Tr}\left(\Pi^2\right)-\mu\left[\text{Tr}\left(k'\Pi\right)\right]^2}{\sqrt{\Omega}},\\
\text{Tr}(\dot kk')=&\frac{\text{Tr}\left(k'\Pi\right)\left[1-\mu\text{Tr}\left(k'^2\right)\right]}{\sqrt{\Omega}},\\
\text{Tr}(\dot k\dot k)=&\frac{\text{Tr}\left(\Pi^2\right)-2\mu\left[\text{Tr}\left(k'\Pi\right)\right]^2+\mu^2\left[\text{Tr}\left(k'\Pi\right)\right]^2\text{Tr}\left(k'^2\right)}{\Omega}.
\end{align}
One can express the $\Pi$-dependent quantities in terms of $k$-dependent quantities by solving these equations above. The solutions show
\begin{align}
\text{Tr}(\dot k\Pi)=&\frac{\text{Tr}(\dot k^2)+\mu\left[\left(\text{Tr}(\dot kk')\right)^2-\text{Tr}(\dot k^2)\text{Tr}\left(k'^2\right)\right]}{\sqrt{1+\mu\left[\text{Tr}(\dot k^2)-\text{Tr}\left(k'^2\right)\right]+\mu^2\left[\left(\text{Tr}(\dot kk')\right)^2-\text{Tr}(\dot k^2)\text{Tr}(k'^2)\right]}},\\
\text{Tr}\left(k'\Pi\right)=&\frac{\text{Tr}(\dot kk')}{\sqrt{1+\mu\left[\text{Tr}(\dot k^2)-\text{Tr}(k'^2)\right]+\mu^2\left[\left(\text{Tr}(\dot kk')\right)^2-\text{Tr}(\dot k^2)\text{Tr}(k'^2)\right]}},\\
\text{Tr}\left(\Pi^2\right)=&\frac{\text{Tr}(\dot k^2)+\text{Tr}(k'^2)+\mu\left[2\left(\text{Tr}(\dot kk')\right)^2-\text{Tr}(\dot k^2)\text{Tr}(k'^2)-\left(\text{Tr}\left(k'^2\right)\right)^2\right]}{1+\mu\left[\text{Tr}(\dot k^2)-\text{Tr}(k'^2)\right]+\mu^2\left[\left(\text{Tr}(\dot kk')\right)^2-\text{Tr}(\dot k^2)\text{Tr}(k'^2)\right]}-\text{Tr}\left(k'^2\right).
\end{align}
Substituting these relations back into the action \eqref{action_pi_k}, we arrive at an action depending on $k$ only 
\begin{align}
\label{action_k}
S[k]=&\frac{\kappa}{4\pi}\int_{\partial M}\frac{1}{\mu}\left(\sqrt{1+\mu\left[\text{Tr}(\dot k^2)-\text{Tr}\left(k'^2\right)\right]+\mu^2\left[\left(\text{Tr}(\dot kk')\right)^2-\text{Tr}(\dot k^2)\text{Tr}\left(k'^2\right)\right]}-1\right)\nonumber\\
&-\Gamma[K].
\end{align}
\par
In the light cone coordinates, this action finally becomes
\begin{align}
\label{action_k}
S[k]=\frac{\kappa}{4\pi}\int_{\partial M}\frac{1}{\mu}\left(\sqrt{1+4\mu \eta^{ij}\mathcal {X}_{ij}+4\mu^2\varepsilon^{ij}\varepsilon^{kl}\mathcal{X}_{ik}\mathcal{X}_{jl}}-1\right)-\Gamma[K]
\end{align}
where $\mathcal{X}_{ij}$ is defined by 
\begin{align}
\mathcal {X}_{ij}=\text{Tr}\left(k^{-1}\partial_ikk^{-1}\partial_{j}k\right),\quad i,j=(+,-),\quad \varepsilon^{+-}=-\varepsilon^{-+}=1.
\end{align}
This is exactly the action for the $T\bar T$-deformed non-chiral WZW model, which is first derived from $T\bar{T}$ flow equation in ~\cite{Bonelli:2018kik}. Therefore, we have verified that the equivalence between the sum of two chiral WZW models and the standard non-chiral WZW model still holds under the $T\bar T$ deformation.
\subsection{Constraints on the $T\bar T$-deformed WZW model}
This mixed boundary condition also gives constraints on the $T\bar T$-deformed WZW model. In order to study the constraints, we consider the Gauss decomposition of $SL(2,\mathbb R)$
\begin{align}
\label{G decomposition}
G=&\left(
  \begin{matrix}
   1 &\  0 \\
   F &\  1 
  \end{matrix}
  \right)
  \left(
  \begin{matrix}
   e^{\phi} &\  0 \\
   0 &\  e^{-\phi} 
  \end{matrix}
  \right)
  \left(
  \begin{matrix}
   1 &\  \Psi \\
   0 &\  1 
  \end{matrix}
  \right),\\
\label{Gbar decomposition}  
\bar{G}=&\left(
  \begin{matrix}
   1 &\  -\bar F \\
   0 &\  1 
  \end{matrix}
  \right)
  \left(
  \begin{matrix}
   e^{-\bar\phi} &\  0 \\
   0 &\  e^{\bar\phi} 
  \end{matrix}
  \right)
  \left(
  \begin{matrix}
   1 &\  0 \\
   -\bar\Psi &\  1 
  \end{matrix}
  \right).
\end{align}
Then the gauge fields $\tilde A ,\bar{\tilde A}$ can be expressed as
\begin{align}
\label{A decomposition}
\tilde A=G^{-1}\text{d}G=\left(
  \begin{matrix}
   \tilde A^0 &\ \tilde A^{-}  \\
   \tilde A^{+} &\ -\tilde A^0 
  \end{matrix}
  \right)=&\left(
  \begin{matrix}
   -e^{2\phi}\Psi\text{d}F+\text{d}\phi &\  -e^{2\phi}\Psi^2\text{d}F+2\Psi\text{d}\phi+\text{d}\Psi \\
   e^{2\phi}\text{d}F&\  e^{2\phi}\Psi\text{d}F-\text{d}\phi 
  \end{matrix}
  \right),\\
\label{Abar decomposition}
\bar{\tilde A}=\bar G^{-1}\text{d}\bar G=\left(
  \begin{matrix}
   \bar{\tilde A}^0 &\ \bar{\tilde A}^{-}  \\
   \bar{\tilde A}^{+} &\ -\bar{\tilde A}^0 
  \end{matrix}
  \right)=&\left(
  \begin{matrix}
   e^{2\bar\phi}\bar\Psi\text{d}\bar F-\text{d}\bar\phi &\ -e^{2\bar\phi}\text{d}\bar F \\
   e^{2\bar\phi}\bar\Psi^2\text{d}\bar F-2\bar\Psi\text{d}\bar\phi-\text{d}\bar\Psi&\  -e^{2\bar\phi}\bar\Psi\text{d}\bar F+\text{d}\bar\phi 
  \end{matrix}
  \right).
\end{align} 
Comparing with ~\eqref{TT deformed A} and~\eqref{TT deformed Abar}, we see that the fields are fixed at $r\to \infty$ as follows:
\begin{align}
\label{TT constraints 1}
e^{2\phi}\partial_{\tilde \theta}F=&\eta r,\quad \partial_{\tilde\theta}\phi=e^{2\phi}\Psi\partial_{\tilde \theta}F,\\
\label{TT constraints 2}
e^{2\bar\phi}\partial_{\tilde \theta}\bar F=&\bar\eta r,\quad \partial_{\tilde\theta}\bar\phi=e^{2\bar\phi}\bar\Psi\partial_{\tilde \theta}\bar F, 
\end{align}
where the parameters $\eta,\bar\eta$ take the form
\begin{align}
\eta=\frac{1-\mu\mathcal{\bar L}_{\mu}}{1-\mu^2\mathcal{L}_{\mu}\mathcal{\bar L}_{\mu}}=&\frac{1}{2}\left[1+\mu(X_{\tilde\theta\tilde\theta}-\bar X_{\tilde\theta\tilde\theta})+\sqrt{1-2\mu(X_{\tilde\theta\tilde\theta}+\bar X_{\tilde\theta\tilde\theta})+\mu^2(X_{\tilde\theta\tilde\theta}-\bar X_{\tilde\theta\tilde\theta})^2}\right],\\
\bar\eta=\frac{1-\mu\mathcal{L}_{\mu}}{1-\mu^2\mathcal{L}_{\mu}\mathcal{\bar L}_{\mu}}=&\frac{1}{2}\left[1-\mu(X_{\tilde\theta\tilde\theta}-\bar X_{\tilde\theta\tilde\theta})+\sqrt{1-2\mu(X_{\tilde\theta\tilde\theta}+\bar X_{\tilde\theta\tilde\theta})+\mu^2(X_{\tilde\theta\tilde\theta}-\bar X_{\tilde\theta\tilde\theta})^2}\right].
\end{align} 
It is useful to write $X_{\tilde\theta\tilde\theta},\bar X_{\tilde\theta\tilde\theta}$ in terms of the parameters 
\begin{align}
\label{X in eta}
X_{\tilde\theta\tilde\theta}=\frac{1}{\mu}\eta(1-\bar\eta),\quad \bar X_{\tilde\theta\tilde\theta}=\frac{1}{\mu}\bar\eta(1-\eta).
\end{align}
According to the constraints~\eqref{TT constraints 1} and~\eqref{TT constraints 2}, we express $\phi',\dot\phi$ and $\Psi',\dot\Psi$ as
\begin{align}
\label{phi-F}
\phi'=&\frac{1}{2}\left(\frac{\eta'}{\eta}-\frac{F''}{F'}\right),\quad 
\dot\phi=\frac{1}{2}\left(\frac{\dot\eta}{\eta}-\frac{\ddot F}{\dot F}\right),\\
\label{pPsi-F}
\Psi'=&\frac{1}{2r}\left(\frac{\eta''}{\eta^2}-\frac{2\eta'^2}{\eta^3}-\frac{F'''}{\eta F'}+\frac{\eta' F''}{\eta^2F'}+\frac{F''^2}{\eta F'^2}\right),\\
\label{dPsi-F}
\dot\Psi=&\frac{1}{2r}\left(\frac{\dot\eta'}{\eta^2}-\frac{2\eta'\dot\eta}{\eta^3}-\frac{\dot F''}{\eta F'}+\frac{\dot\eta F''}{\eta^2F'}+\frac{F''\dot F'}{\eta F'^2}\right),
\end{align} 
where the overdot and prime denote the derivative with respect to $\tilde t$ and $\tilde \theta$. Similar relations for the $\bar\phi',\dot{\bar\phi}$ and $\bar\Psi',\dot{\bar\Psi}$ can also be obtained. For the Brown-Henneaux boundary condition, the parameters $\eta,\bar\eta$ are both equal to $1$. Then, the constraints can reduce the WZW model to Alekseev-Shatashvili action. However, when the deformation is turned on, the parameters $\eta,\bar\eta$ appear in the constraints. In order to make a further reduction, we have to find the relations between $\eta,\bar\eta$ and $F,\bar F$.
\par
In fact, one can rewrite $X_{\tilde\theta\tilde\theta}$ and $\bar X_{\tilde\theta\tilde\theta}$ in Gauss parametrization. As a consequence, \eqref{X in eta} implies the differential equations for $\eta$ and $\bar\eta$ 
\begin{align}
\label{eta-equation}
\frac{\eta''}{\eta}-\frac{3}{2}\left(\frac{\eta'}{\eta}\right)^2-\{F;\tilde\theta\}=\frac{1}{\mu}\eta(1-\bar\eta),\\
\label{etabar-equation}
\frac{\bar\eta''}{\bar \eta}-\frac{3}{2}\left(\frac{\bar \eta'}{\bar\eta}\right)^2-\{\bar F;\tilde\theta\}=\frac{1}{\mu}\bar\eta(1-\eta),
\end{align}
where $\{f;\tilde\theta\}$ represents Schwarzian derivative defined by
\begin{align}
\{f;\tilde\theta\}=\frac{f'''}{f'}-\frac{3}{2}\left(\frac{f''}{f'}\right)^2.
\end{align}
Although it is difficult to get the exact solutions, we can find the perturbation solutions in the first few orders of small $\mu$
\begin{align}
\label{perturbation solution}
\eta=1+\mu \{\bar F;\tilde\theta\}+O(\mu^2),\quad \bar\eta=1+\mu \{F;\tilde\theta\}+O(\mu^2).
\end{align}   
\par
In the Gauss parametrization, we can reduce the $T\bar{T}$-deformed WZW model into
\begin{align}
\label{constrained TT deformed chiral WZW action}
S_{\text{total}}=&\frac{\kappa}{4\pi}\int_{\partial M}d\tilde\theta d\tilde t\left(\frac{\dot{\eta }'}{\eta }-\frac{3 \dot{\eta } \eta '}{2 \eta ^2}-\frac{\eta '\dot{F}'}{2 \eta  F'}+\frac{\dot{\eta } F''}{2 \eta  F'}-\frac{\dot{F}''}{F'}+\frac{3 \dot{F}' F''}{2 F'^2}\right)\nonumber\\
-&\frac{\kappa}{4\pi}\int_{\partial M}d\tilde\theta d\tilde t\left(\frac{\dot{\bar \eta }'}{\bar \eta }-\frac{3 \dot{\bar \eta } \bar \eta '}{2 \bar \eta ^2}-\frac{ \bar \eta '\dot{\bar F}'}{2 \bar \eta  \bar F'}+\frac{\dot{\bar \eta } \bar F''}{2 \bar \eta  \bar F'}-\frac{\dot{\bar F}''}{\bar F'}+\frac{3 \dot{\bar F}' \bar F''}{2 \bar F'^2}\right)\nonumber\\
+&\frac{\kappa}{4\pi\mu}\int_{\partial M}d\tilde\theta d\tilde t(\eta+\bar\eta-2),
\end{align}
where $\eta,\bar\eta$ are determined by the equations~\eqref{eta-equation} and~\eqref{etabar-equation}. Moreover, it is useful to parametrize the boundary value of $F$ and $\bar{F}$ as
\begin{align}
F=\tan\left(\frac{\xi}{2}\right),\quad \bar F=\tan\left(\frac{\bar\xi}{2}\right),
\end{align} 
such that $\xi,\bar\xi$ are valued in $\text{Diff}(S^1)/PSL(2,\mathbb R)$~\cite{Cotler:2018zff}. Then we find the relations 
\begin{align}
\label{F-xi}
\frac{\dot{F}''}{F'}-\frac{3 \dot{F}' F''}{2 F'^2}=\frac{d}{d\tilde t}\left(\frac{\xi''}{\xi'}\right)+\frac{1}{2}\left(\xi'\dot\xi-\frac{\xi''\dot\xi'}{\xi'^2}\right),\\
\label{Fbar-xibar}
\{F;\tilde\theta\}=\{\xi;\tilde\theta\}+\frac{1}{2}\xi'^2=\frac{d}{d\tilde\theta}\left(\frac{\xi''}{\xi'}\right)+\frac{1}{2}\left(\xi'^2-\frac{\xi''^2}{\xi'^2}\right),
\end{align}  
as well as the similar relations for the barred quantities. In order to see whether the resulting theory is a $T\bar{T}$-deformed conformal theory, we can consider the perturbation form of this action. Plugging~\eqref{perturbation solution} into the action~\eqref{constrained TT deformed chiral WZW action} and dropping some total derivative terms, we finally arrive at
\begin{align}
\label{TT-deformed AS action}
S_{\text{total}}=&-\frac{\kappa}{8\pi}\int_{\partial M}d\tilde\theta d\tilde t\left[\left(\frac{\xi''\partial_{-}\xi'}{\xi'^2}-\xi'\partial_{-}\xi\right)-\left(\frac{{\bar\xi}''\partial_{+}{\bar\xi}'}{{\bar\xi}'^2}-{\bar\xi}'\partial_{+}{\bar\xi}\right)\right]\nonumber\\
&+\frac{\mu\kappa}{16\pi}\int_{\partial M}d\tilde\theta d\tilde t\left[\left(\{\xi;\tilde\theta\}+\frac{1}{2}\xi'^2\right)\left(\{\bar\xi;\tilde\theta\}+\frac{1}{2}\bar\xi'^2\right)\right]+O(\mu^2).
\end{align} 
The leading order is exactly the sum of left-moving and right-moving Alekseev-Shatashvili quantization of coadjoint orbit $\text{Diff}(S^1)/PSL(2,\mathbb R)$ of the Virasoro group~\cite{Cotler:2018zff, Witten:1987ty, Alekseev:1988ce}. The first order correction is nothing but coupling these two copies through the $T\bar{T}$ deformation, since the stress tensors of chiral Alekseev-Shatashvili actions are exactly given by  
\begin{align}
T_{\text{L}}=\{\xi;\tilde\theta\}+\frac{1}{2}\xi'^2,\quad \bar T_{\text{R}}=\{\bar\xi;\tilde\theta\}+\frac{1}{2}\bar\xi'^2.
\end{align}
Therefore, the boundary dynamics of AdS$_3$ with mixed boundary condition is described by the action~\eqref{TT-deformed AS action}, which is a $T\bar T$-deformed conformal theory in first order as expected. In~\cite{Ouyang:2020rpq}, very similar results were obtained from a boundary WZW model through the $T\bar{T}$ flow. These results may give a precise check on the correspondence between the $T\bar{T}$-deformed CFT and AdS$_3$ gravity with the mixed boundary condition.  
\section{$J\bar T$ deformation}
\label{JT deformation}
Another interesting integrable deformation is the $J\bar{T}$ deformation~\cite{Guica:2017lia}. In this section, we would like to study the $J\bar{T}$ deformation. We firstly give a brief review for the boundary condition for $J\bar T$-deformed CFT. In Chern-Simons form, this boundary condition implies a certain nontrivial boundary term. The spectrum of $J\bar T$-deformed CFT is obtained from this boundary term in the Hamiltonian form. We will also show that the asymptotic boundary dynamics is described by one type of the $J\bar T$-deformed chiral WZW model.      
\subsection{Review of the boundary condition for the $J\bar{T}$ deformation}
By the definition of $J\bar T$ deformation, its action could be written as 
\begin{align}
\label{JT action}
\frac{\partial}{\partial \mu} S_{J\bar{T}}=\int d^2x\sqrt{\gamma}\varepsilon^{ij}J_{i}T_{j\bar z}=\int d^2xe\varepsilon^{ij}J_{i}T^a_{j}e_{a\bar z}.
\end{align}
For convenience, we have written it in vielbein form. In this model, we have to consider the CFT involving stress tensor $T^a_i$ and the conserved current $J^i$, which are canonically conjugate to the boundary vielbein $e^i_a$ and the gauge field $\Phi_i$. Then the variation of the original CFT action would be 
\begin{align}
\delta S_{CFT}=\int d^2xe\left(T_i^a\delta e^i_a+J^i\delta \Phi_i\right). 
\end{align}
When the deformation is turned on, we may suppose the variation takes the following form 
\begin{align}
\delta S_{J\bar{T}}=\int d^2x\tilde e\left(\tilde T_i^a\delta \tilde e^i_a+\tilde J^i\delta \tilde \Phi_i\right). 
\end{align}
The deformed quantities are marked with a tilde. In~\cite{Bzowski:2018pcy}, by using the $J\bar{T}$ flow equation~\eqref{JT action}, the $J\bar T$-deformed variables were constructed from the original theory
\begin{align}
\label{deformed variables 1}
\tilde{e}^i_a=&e^i_a-\mu_aJ^i,\quad \tilde{\Phi}_i=\Phi_i-\mu_aT^a_i,\\
\label{deformed variables 2}
\tilde T^a_i=&T^a_i+(\mu_bT^b_{j}J^j)(e^a_i+\mu_{i}J^a),\quad \tilde J^i=J^i.
\end{align}
We focus mainly on the deformed vielbein $\tilde{e}^i_a$ and the gauge field $\tilde{\Phi}_i$, which could help us to fix the boundary condition of AdS$_3$. 
\par
On the gravity side, we have to introduce a $U(1)$ Chern-Simons gauge field coupling with  AdS$_3$ gravity. Therefore, the total action associated with the $J\bar T$ deformation should be 
\begin{align*}
S_{\rm{total}}=&S_{\rm{grav}}+S_{\rm{U(1)}}
=\int_M d^3x\sqrt{g}\left[\frac{1}{16\pi G}\left(R+\frac{2}{l^2}\right)+\frac{\kappa'}{4\pi}\varepsilon^{\mu\nu\rho}\Phi_{\mu}\partial_{\nu}\Phi_{\rho} \right],
\end{align*}
where $k'$ is the $U(1)$ Chern-Simons level. Generally, the $U(1)$ charge is introduced by adding a Maxwell term, such as the charged black hole. Since we are working in an odd-dimensional spacetime, this gauge field have the $U(1)$ Chern-Simons form.  In order to ensure the variational process, we add the Gibbons-Hawking boundary term for the gravitational part. As for the gauge field part, the boundary term turns out to be 
\begin{align}
S_{\rm{U(1)-bdy}}=\frac{\kappa'}{8\pi}\int_{\partial M} d^2x\sqrt{\gamma}\gamma^{ij}\Phi_{i}\Phi_{j},
\end{align}
where $\gamma_{ij}$ is the induced metric on the boundary $\partial M$. Then the variation of total action in the bulk becomes 
\begin{align}
\delta S_{\rm{total}}=-\frac{1}{2}\int_{\partial M}d^2x\sqrt{\gamma}\left(T^{\rm{grav}}_{ij}+T^{\rm{U(1)}}_{ij}\right)\delta \gamma^{ij}-\int_{\partial M}d^2x\sqrt{\gamma}J^i\delta \Phi_{i},
\end{align}
with
\begin{align}
\label{BY stress tensor}
T^{\rm{grav}}_{ij}=&\frac{1}{8\pi G}\left(K_{ij}-\gamma_{ij}K+\gamma_{ij}\right),\\
\label{CS stress tensor}
T^{\rm{U(1)}}_{ij}=&\frac{\kappa'}{4\pi}\left(\Phi_i\Phi_j-\frac{1}{2}\gamma_{ij}\Phi^2\right),\\
\label{current}
J^{i}=&\frac{\kappa'}{4\pi}(\gamma^{ij}-\varepsilon^{ij})\Phi_{j}.
\end{align} 
where $T^{\rm{grav}}_{ij}$ is the Brown-York stress tensor~\cite{Brown:1992br, Brown:1994gs}, $T^{\rm{U(1)}}_{ij}$ comes from the $U(1)$ Chern-Simons boundary term and $J^i$ is the $U(1)$ conserved current. This is the basic structure in AdS$_3$/CFT$_2$ correspondence with additional $U(1)$ charge~\cite{Kraus:2006wn}. 
\par
In Fefferman-Graham gauge, the deformed vielbein~\eqref{deformed variables 1} corresponds to fixing the $g^{(0)}_{ij}$ as 
\begin{align} 
g^{(0)}_{++}=-\mu J(x^{+}),\quad g^{(0)}_{-+}=g^{(0)}_{+-}=\frac{1}{2},\quad g^{(0)}_{--}=0,
\end{align}
which can be obtained from the Ba\~nados geometry through a coordinate transformation
\begin{align}
\label{JT coordinate transformation}
dz=dx^{+}, \quad  d\bar z=dx^{-}-\mu J(x^{+})dx^{+}.
\end{align}
Therefore, the deformed solution is parametrized by $\mathcal L_{\mu},\mathcal {\bar L}_{\mu}, J$
\begin{align} 
\mathcal L_{\mu}=\mathcal L(x^{+}),\quad  \mathcal {\bar L}_{\mu}=\mathcal L(x^{-}-\mu\int J(x^{+})dx^{+}),\quad J=J(x^{+}).
\end{align}
We use similar notations for the $J\bar T$ deformation, these notations should not be confused with the $T\bar T$ deformation. A very similar boundary condition for AdS$_3$ has been considered in~\cite{Compere:2013bya}, when they studied $SL(2, \mathbb R)\times U(1)$ symmetries in AdS$_3$.
\par
In addition, we also need to fix the gauge field $\tilde\Phi$. From~\eqref{current}, the gauge field $\tilde\Phi$ can be written as 
\begin{align}
\label{gauge field 1} 
\tilde\Phi_{-}=&\mathcal F(x^{-},x^{+}),\\
\label{gauge field 2} 
\tilde\Phi_{+}=&\frac{2\pi}{k}J(x^{+})-\mu J(x^{+})\mathcal F(x^{+},x^{-}).
\end{align}
Comparing the deformed gauge field $\tilde\Phi$ with \eqref{deformed variables 1}, we can identify 
\begin{align} 
\mathcal F=\mu T_{--},\quad -\mu J(x^{+})\mathcal F=\mu T_{-+},
\end{align}
where $T_{ij}$ is the total stress tensor of the system 
\begin{align} 
T_{ij}=T^{\rm{grav}}_{ij}+T^{\rm{CS}}_{ij}.
\end{align}
This means that the additional boundary term of the $U(1)$ Chern-Simons action have a backreaction for the formalism of deformed gauge field. Finally, one arrives at the equation for $\mathcal F$  
\begin{align}
\label{gauge field F} 
\mathcal F=&\frac{\kappa \mu}{2\pi}\mathcal{\bar L}_{\mu}+\frac{\mu\kappa'}{4\pi}\mathcal F^2,\\
\text{or}\qquad \mathcal F=&\frac{2\pi}{\mu\kappa'}\left(1-\sqrt{1-\frac{\mu^2\kappa\kappa'}{2\pi^2}\mathcal{\bar L}_{\mu}}\right).
\end{align}
\par
We summarize the mixed boundary conditions to complete this subsection. The mixed boundary condition for $J\bar{T}$ deformation includes fixing AdS$_3$ metric as well as  $U(1)$ gauge field. The AdS$_3$ metric is determined by a coordinate transformation~\eqref{JT coordinate transformation}. The gauge field refers to the stress tensor of the whole system through \eqref{gauge field 1} and \eqref{gauge field F}. As a result, we can express the metric and gauge field in terms of $\mathcal L,\mathcal {F}, J$. Moreover, this mixed boundary condition would imply the asymptotic dynamics because it is defined at infinity.
\subsection{Chern-Simons formalism and the boundary term}
Now we put the mixed boundary condition in the Chern-Simons formalism to find out the associated  boundary term. As mentioned above, the total action in the bulk consists of the gravitational part and the $U(1)$ Chern-Simons gauge field part. For the gravitational part, the action can be formulated in $SL(2,\mathbb R)\times SL(2,\mathbb R)$ Chern-Simons theory. Therefore, the total action would be 
\begin{align}
\label{CS with U}
S(\tilde A,\bar{\tilde A},\tilde \Phi)=I(\tilde A)-I(\bar {\tilde A})+\frac{\kappa'}{4\pi}\int_{M}\tilde\Phi\wedge \text{d}\tilde \Phi.
\end{align} 
By using the coordinate transformation~\eqref{JT coordinate transformation}, we obtain the $SL(2,\mathbb{R})$ gauge fields
\begin{align}
\label{JT deformed A}
\tilde A=&\frac{1}{r}L_0dr+\left(rL_{-1}+\frac{1}{r}\mathcal {L}L_1\right)dx^{+},\\
\label{JT deformed Abar}
\bar{\tilde A}=&-\frac{1}{r}L_0dr+\left(\frac{1}{r}\mathcal{ \bar L}_{\mu}L_{-1}+rL_{1}\right)(dx^{-}-\mu J(x^{+})dx^{+}),
\end{align}
which still satisfy the equations of motion. After eliminating the radial coordinates, we write down the induced connections  
\begin{align}
\tilde a=&(L_{-1}+\mathcal L(x^{+})L_{1})dx^{+},\\
\bar{\tilde a}=&(\mathcal{ \bar L}_{\mu}L_{-1}+L_{1})(dx^{-}-\mu J(x^{+})dx^{+}).
\end{align}
Clearly, the left chiral boundary condition is maintained, but the right chiral boundary condition is broken. Besides, the $U(1)$ gauge field $\tilde{\Phi}$ is fixed in~\eqref{gauge field 1} and~\eqref{gauge field 2}. In the coordinates $\tilde\theta=(x^++x^-)/2,\tilde t=(x^+-x^-)/2$, the mixed boundary condition becomes 
\begin{align}
\label{a-jt}
\tilde a_{\tilde{\theta}}=&L_{-1}+\mathcal L(x^{+})L_{1},\quad \tilde a_{\tilde{t}}=\tilde a_{\tilde{\theta}},\\ 
\label{abar-jt}
\bar{\tilde a}_{\tilde\theta}=&(\mathcal{\bar L}_{\mu}L_{-1}+L_{1})(1-\mu J),\quad \bar{\tilde a}_{\tilde t}=-\frac{1+\mu J}{1-\mu J}\bar{\tilde a}_{\tilde\theta},\\
\label{phi-jt}
\tilde\Phi_{\tilde \theta}=&\frac{2\pi}{\kappa'}J+(1-\mu J)\mathcal F,\quad \tilde\Phi_{\tilde t}=\frac{4\pi}{\kappa'}\frac{J}{1-\mu J}-\frac{1+\mu J}{1-\mu J}\tilde \Phi_{\tilde \theta}.
\end{align}
This boundary condition requires a boundary term be added to the action~\eqref{CS with U}, which turns out to be 
\begin{align}
\label{JT boundary term}
\mathcal B=-\frac{\kappa}{4\pi }\int_{\partial M}d\tilde td\tilde\theta\left[\mathcal L-\frac{2\pi^2}{\kappa\kappa'}J^2+\frac{2\pi}{\mu\kappa}(1-\mu J)\mathcal F\right].
\end{align}
The detailed derivation of this boundary term is given in Appendix~\ref{variation JTbar}. This boundary term also reduces to the CFT case when $\mu\to 0$. In addition, it provides a coupling between the right chiral Chern-Simons theory and a $U(1)$ gauge field, but keeps the left chiral Chern-Simons action unchanged. 
\par
In the Hamiltonian form, this boundary term gives the surface integral 
\begin{align}
E=&\frac{\kappa}{4\pi}\int d\tilde\theta\left[\mathcal L-\frac{2\pi^2}{\kappa\kappa'}J^2+\frac{4\pi^2}{\mu^2\kappa\kappa'}(1-\mu J)\left(1-\sqrt{1-\frac{\mu^2\kappa\kappa'}{2\pi^2}\mathcal {\bar L}_{\mu}}\right)\right].
\end{align}
We consider the BTZ black holes, in which $\mathcal L$ and $\mathcal{\bar L}$ are constants. After rescaling the coordinates~\cite{Bzowski:2018pcy}, we can identify
\begin{align}
\mathcal L=&\frac{16\pi^2 G(\Delta-c/24)}{R^2}=\frac{4\pi^2(\Delta-c/24)}{\kappa R^2},\quad J=\frac{Q_0}{R},\\
\mathcal{\bar L}_{\mu}=&\frac{\mathcal {\bar L}}{(1-\mu J)^2}=\frac{16\pi^2G(\bar\Delta-c/24)}{R^2(1-\mu Q_0/R)^2}=\frac{4\pi^2(\bar\Delta-c/24)}{\kappa R^2(1-\mu Q_0/R)^2}.
\end{align}
Up to a coefficient, the surface integral ends up with 
\begin{align}
E=&\frac{2\pi(\Delta-c/24)}{R}-\frac{2\pi}{\kappa'}\frac{Q_0^2}{R}+\frac{4\pi}{\mu^2\kappa'}(R-\mu Q_0)\left(1-\sqrt{1-\frac{2\mu^2\kappa'(\bar\Delta-c/24)}{ (R-\mu Q_0)^2}}\right).
\end{align}
which is the spectrum of the $J\bar{T}$ deformed CFT in~\cite{Bzowski:2018pcy, Chakraborty:2018vja}, as expected. Here we reproduce the spectrum from gravity side using the surface integral method. Just as in the case of $T\bar T$ deformation, the boundary term is defined at infinity. From the holographic point of view, the $J\bar{T}$ deformation corresponds actually to a deformation of the boundary condition of AdS$_3$, which can be treated as a coordinate transformation. This asymptotic boundary condition may imply the boundary dynamics, and we would like to discuss this in later subsections.
\subsection{From Chern-Simons theory to $J\bar{T}$-deformed WZW model}
We then follow the method used in $T\bar T$ deformation to study the asymptotic dynamics for this mixed boundary condition. By using~\eqref{a-jt},~\eqref{abar-jt} and~\eqref{phi-jt}, one gets 
\begin{align}
J=&\frac{1}{\mu}\left(1-\sqrt{\left(1-\frac{\mu\kappa'}{2\pi}\tilde\Phi_{\tilde\theta}\right)^2+\frac{\mu^2\kappa\kappa'}{4\pi^2}\bar X_{\tilde\theta\tilde\theta}}\right),\\
\mathcal F=&\frac{\tilde\Phi_{\tilde\theta}-{2\pi J}/{\kappa'}}{1-\mu J}.
\end{align}
Plugging into~\eqref{JT boundary term}, the boundary term becomes
\begin{align}
\label{JT total boundary term}
\mathcal B=&-\int d\tilde t d\tilde\theta\left[\frac{\kappa}{4\pi}X_{\tilde\theta\tilde\theta}+\frac{\kappa'}{4\pi}\tilde\Phi_{\tilde\theta}^2+\frac{\kappa}{4\pi}\bar X_{\tilde\theta\tilde\theta}\right]\nonumber\\
&+\int d\tilde t d\tilde\theta\frac{2\pi}{\mu^2\kappa'}\left(1-\frac{\mu\kappa'}{2\pi}\tilde\Phi_{\tilde\theta}-\sqrt{\left(1-\frac{\mu\kappa'}{2\pi}\tilde \Phi_{\tilde \theta}\right)^2+\frac{\mu^2\kappa\kappa'}{4\pi^2}\bar X_{\tilde\theta\tilde\theta}}\right).
\end{align}
Finally, the total Chern-Simons action with this certain boundary term can be reduced to 
\begin{align}
\label{JT deformed WZW action}
S_{\rm{total}}
=&\frac{\kappa}{4\pi}\int d\tilde t d\tilde\theta\left(X_{\tilde\theta\tilde t}- X_{\tilde\theta\tilde\theta}-\bar X_{\tilde\theta\tilde t}-\bar X_{\tilde\theta\tilde\theta}\right)+\Gamma[g]-\Gamma[\bar g]\nonumber\\
&+\frac{\kappa'}{4\pi}\int d\tilde t d\tilde\theta\left(\tilde\Phi_{\tilde\theta}\tilde\Phi_{\tilde t}-\tilde\Phi_{\tilde\theta}^2\right)\nonumber\\
&+\frac{2\pi}{\mu^2\kappa'}\int d\tilde t d\tilde\theta\left(1-\frac{\mu\kappa'}{2\pi}\tilde\Phi_{\tilde\theta}-\sqrt{\left(1-\frac{\mu\kappa'}{2\pi}\tilde \Phi_{\tilde \theta}\right)^2+\frac{\mu^2\kappa\kappa'}{4\pi^2}\bar X_{\tilde\theta\tilde\theta}}\right).
\end{align}
This is actually one type of the $J\bar T$-deformed WZW action, which can also be got from $J\bar T$ flow equation by adding an extra $U(1)$ gauge field, see Appendix~\ref{JT deformed WZW model} for details. The effect of  $J\bar T$ deformation is coupling the right-moving $SL(2,\mathbb R)$ WZW model with left-moving $U(1)$ gauge field. From the perspective of holography, the boundary dynamics of AdS$_3$ with the mixed boundary condition can be described by~\eqref{JT deformed WZW action}, namely a $J\bar{T}$-deformed conformal theory.
\par
We give some comments about the $J\bar T$-deformed WZW model. The difference between the $J\bar T$-deformed scalar field and the $J\bar T$-deformed WZW model is the definition of $U(1)$ current $J$. In the latter one, the current $J$ is introduced through adding an extra $U(1)$ gauge field. Of course, one can do the deformation by using one component of $SL(2,\mathbb R)$ current $J^a$, such as $J^{0}$. However, there will be another boundary condition for AdS$_3$ instead of the mixed one. We will not discuss this case in this paper. 
\subsection{Constraints on the $J\bar{T}$-deformed WZW model} 
We now consider constraints on the $J\bar T$-deformed WZW model. We will use the same notation as in the $T\bar{T}$ deformation. By using the Gauss decomposition~\eqref{A decomposition} and \eqref{Abar decomposition}, the boundary condition~\eqref{JT deformed A} and~\eqref{JT deformed Abar} imply the constraints
\begin{align}
\label{JT constraints 1}
e^{2\phi}\partial_{\tilde \theta}F=r,\quad \partial_{\tilde\theta}\phi=e^{2\phi}\Psi\partial_{\tilde \theta}F,\\
\label{JT constraints 2}
e^{2\bar\phi}\partial_{\tilde \theta}\bar F=\bar\zeta r,\quad \partial_{\tilde\theta}\bar\phi=e^{2\bar\phi}\bar\Psi\partial_{\tilde \theta}\bar F,
\end{align}
where
\begin{align}
\label{zeta to X}
\bar\zeta=&(1-\mu J)=\sqrt{\left(1-\frac{\mu\kappa'}{2\pi}\tilde\Phi_{\tilde\theta}\right)^2+\frac{\mu^2\kappa\kappa'}{4\pi^2}\bar X_{\tilde\theta\tilde\theta}},\nonumber\\
\text{or}\quad \bar X_{\tilde\theta\tilde\theta}=&\frac{4\pi^2}{\mu^2\kappa\kappa'}\left[{\bar\zeta}^2-\left(1-\frac{\mu\kappa'}{2\pi}\tilde\Phi_{\tilde\theta}\right)^2\right].
\end{align}
The left-moving part remains unchanged, but the right-moving part is deformed because of $\bar\zeta\ne 1$. From these constraints, one can express $\phi',\dot{\phi}$ and $\Psi',\dot{\Psi}$ in terms of $F$
\begin{align}
\phi'=&-\frac{F''}{2F'},\quad \dot\phi=-\frac{\dot F'}{2F'},\\
\Psi'=&\frac{1}{2r}\left(-\frac{F'''}{F'}+\frac{F''^2}{F'^2}\right),\quad 
\dot\Psi=\frac{1}{2r}\left(-\frac{\dot F''}{F'}+\frac{F''\dot F'}{ F'^2}\right).
\end{align} 
Similarly we have
\begin{align}
\bar\phi'=&\frac{1}{2}\left(\frac{{\bar\zeta}'}{{\bar\zeta}}-\frac{{\bar F}''}{{\bar F}'}\right),\quad 
\dot{\bar\phi}=\frac{1}{2}\left(\frac{\dot{\bar\zeta}}{{\bar\zeta}}-\frac{\dot{\bar F}'}{{\bar F}'}\right),\\
\bar{\Psi}'=&\frac{1}{2r}\left(\frac{{\bar\zeta}''}{{\bar\zeta}^2}-\frac{2{\bar\zeta}'^2}{{\bar\zeta}^3}-\frac{{\bar F}'''}{{\bar\zeta} {\bar F}'}+\frac{{\bar\zeta}' {\bar F}''}{{\bar\zeta}^2{\bar F}'}+\frac{{\bar F}''^2}{{\bar\zeta} {\bar F}'^2}\right),\\
\dot{\bar{\Psi}}=&\frac{1}{2r}\left(\frac{\dot{\bar\zeta}'}{{\bar\zeta}^2}-\frac{2{\bar\zeta}'\dot{\bar\zeta}}{{\bar\zeta}^3}-\frac{\dot {\bar F}''}{{\bar\zeta} {\bar F}'}+\frac{\dot{\bar\zeta} {\bar F}''}{{\bar\zeta}^2{\bar F}'}+\frac{{\bar F}''\dot {\bar F}'}{{\bar\zeta} {\bar F}'^2}\right).
\end{align}
According to these relations, we get the differential equation for $\bar{\zeta}$
\begin{align}
\label{equation for zetabar}
\frac{\bar\zeta''}{\bar\zeta}-\frac{3}{2}\left(\frac{\bar\zeta'}{\bar\zeta}\right)^2-\{\bar F;\tilde\theta\}=\frac{4\pi^2}{\mu^2\kappa\kappa'}\left[{\bar\zeta}^2-\left(1-\frac{\mu\kappa'}{2\pi}\tilde\Phi_{\tilde\theta}\right)^2\right].
\end{align} 
The solutions of this equation allow us to express the parameter $\bar\zeta$ in terms of $\bar F$ and $\tilde{\Phi}$. The perturbation solution in the first few orders of small $\mu$ is
\begin{align}
\label{perturbation jt}
\bar\zeta=1-\frac{\mu\kappa'}{2\pi}\tilde\Phi_{\tilde\theta}+\frac{\mu^2\kappa\kappa'}{8\pi^2}\{\bar F;\tilde\theta\}+O(\mu^3),
\end{align} 
which can be used to give a further reduction of the deformed WZW action.
\par
Finally, the total action~\eqref{JT deformed WZW action} can be expressed in  Gauss parametrization 
\begin{align}
\label{constrained JT deformed WZW}
S_{\rm{total}}=&\frac{\kappa}{4\pi}\int d\tilde t d\tilde\theta\left(\{F,\tilde\theta\}+\frac{3F''\dot F'}{2F'^2}-\frac{\dot F''}{F'}\right)+\frac{\kappa'}{4\pi}\int d\tilde t d\tilde\theta\left(\tilde\Phi_{\tilde\theta}\tilde\Phi_{\tilde t}-\tilde\Phi_{\tilde\theta}^2\right)\nonumber\\
-&\frac{\kappa}{4\pi}\int d\tilde t d\tilde\theta\left(\frac{\dot{{\bar\zeta} }'}{{\bar\zeta} }-\frac{3 \dot{{\bar\zeta} } {\bar\zeta} '}{2 {\bar\zeta}^2}-\frac{{\bar\zeta} '\dot{\bar{F}}'}{2{\bar\zeta} \bar{F}'}+\frac{\dot{{\bar\zeta} } \bar{F}''}{2{\bar\zeta} \bar{F}'}-\frac{\dot{\bar{F}}''}{\bar{F}'}+\frac{3 \dot{\bar{F}}' \bar{F}''}{2 \bar{F}'^2}\right)\nonumber\\
-&\frac{\kappa}{4\pi}\int d\tilde t d\tilde\theta\left(\frac{{\bar\zeta} ''}{{\bar\zeta} }-\frac{3{\bar\zeta}'^2}{2{\bar\zeta} ^2}-\{\bar F;\tilde\theta\}\right)+\frac{2\pi}{\mu^2\kappa'}\int d\tilde t d\tilde\theta\left(1-\frac{\mu\kappa'}{2\pi}\tilde\Phi_{\tilde\theta}-\bar\zeta\right).
\end{align}
Again, one can parametrize the $F$ and $\bar F$ to the angular variables $\xi$ and $\bar\xi$. Substituting the perturbation solution~\eqref{perturbation jt} into the action, we arrive at 
\begin{align}
\label{constrained JT deformed WZW-perturbative}
S_{\text{total}}=&-\frac{\kappa}{8\pi}\int_{\partial M}d\tilde\theta d\tilde t\left[\left(\frac{\xi''\partial_{-}\xi'}{\xi'^2}-\xi'\partial_{-}\xi\right)-\left(\frac{{\bar\xi}''\partial_{+}{\bar\xi}'}{{\bar\xi}'^2}-{\bar\xi}'\partial_{+}{\bar\xi}\right)\right]\nonumber\\
&+\frac{\kappa'}{4\pi}\int d\tilde t d\tilde\theta\left(\tilde\Phi_{\tilde\theta}\tilde\Phi_{\tilde t}-\tilde\Phi_{\tilde\theta}^2\right)+\frac{\mu\kappa\kappa'}{8\pi^2}\int_{\partial M}d\tilde\theta d\tilde t\tilde\Phi_{\theta}\left(\{\bar\xi;\tilde\theta\}+\frac{1}{2}\bar\xi'^2\right)+O(\mu^2).
\end{align} 
The leading order of this action is the sum of two opposite chiral Alekseev-Shatashvili actions with an additional $U(1)$ gauge field. The first order correction is just the coupling of the right-moving Alekseev-Shatashvili action and the left-moving $U(1)$ gauge field through the $J\bar{T}$ operator. Consequently, the asymptotic boundary dynamics of AdS$_3$ with this mixed boundary condition is described by one type of $J\bar{T}$-deformed Alekseev-Shatashvili action. 
However, since our construction depends on introduction of a gauge fields $\tilde\Phi$, the resultant theory should differ from the standard $J\bar{T}$ deformation. The latter is the coupling of two opposite chiral Alekseev-Shatashvili actions without additional gauge fields. 
\section{Conclusion and discussion}
\label{Conclusion and discussion}
In this paper, we study the holographic aspects of $T\bar T/J\bar T$-deformed CFTs in Chern-Simons formalism. It is shown that the deformed CFTs correspond to AdS$_3$ with mixed boundary conditions. Based on the mixed boundary condition, the certain boundary terms are obtained. We also show that the boundary dynamics of Chern-Simons AdS$_3$ gravity turns out to be the $T\bar T/J\bar T$-deformed WZW model.  
\par
Unlike the cutoff point of view, the mixed boundary condition for the $T\bar{T}$ deformation is defined at infinity. We find that this boundary condition implies a nontrivial boundary term in Chern-Simons formalism. The boundary term gives rise to total energy of this system, which matches with the spectrum of $T\bar{T}$-deformed CFT. This spectrum is exactly the quasi-local energy of BTZ black hole, if we identify $\mu=1/r_c^2$. After writing the boundary term in terms of gauge fields, the total action can reduce to $T\bar T$-deformed two chiral WZW models. The effect of $T\bar{T}$ deformation is coupling the two chiral WZW models. Moreover, the mixed boundary condition also gives the constraints on $T\bar{T}$-deformed WZW model. By disentangling the constraints, the boundary theory turns out to be the $T\bar{T}$-deformed Alekseev-Shatashvili quantization of coadjoint orbit of the Virasoro group. Finally, we show that the $T\bar T$-deformed standard non-chiral WZW model is equivalent to the $T\bar T$-deformed two chiral WZW models.
\par
As for the $J\bar T$ deformation, the holographic interpretation is also AdS$_3$ gravity but with an extra $U(1)$ Chern-Simons gauge field coupling to the gravity. After rewriting the gravitational action in Chern-Simons formalism, we also obtain the associated boundary term. As expected, this boundary term precisely gives the spectrum of $J\bar T$-deformed CFT. In addition, based on this nontrivial boundary term, the boundary dynamics is also studied. It turns out that the boundary dynamics of AdS$_3$ can be described by one type of constrained $J\bar T$-deformed WZW model. This type of $J\bar{T}$-deformed WZW model can also be obtained from the $J\bar T$ flow equation through adding a supplementary $U(1)$ gauge field. However, this type of $J\bar T$-deformed WZW model turns out to be a coupling of the right-moving Alekseev-Shatashvili action to a $U(1)$ gauge field. The standard $J\bar T$ deformation should be the coupling of two opposite chiral Alekseev-Shatashvili actions via the $J\bar{T}$ operator. Regarding to this, it would be interesting to find another boundary condition in the bulk and perform a holographic check.
\par
Furthermore, we show that the effect of $T\bar T$ deformation is the coupling of two opposite chiral $SL(2, \mathbb R)$ WZW models, and the effect of $J\bar T$ deformation is coupling a right-moving $SL(2, \mathbb R)$ WZW model with a $U(1)$ WZW model. It would be interesting to consider $SL(N, \mathbb R)$ WZW models and couple two WZW models through higher spin currents deformation, since $SL(N, \mathbb R)$ WZW models correspond to higher spin gravity~\cite{Gaberdiel:2010ar, Henneaux:2010xg, Campoleoni:2010zq}. This will be helpful to understand the holographic aspects of higher spin gravity under the integrable deformation.

\section*{Acknowledgments}
We would like to thank Song He for useful discussions. We are grateful to Chen-Te Ma for drawing our attention to Ref.~\cite{Cotler:2018zff}, which inspires us to study the constraints on the deformed WZW models. This work is supported by the National Natural Science Foundation of China (NSFC) with Grants No.11875082 and No.11947302.

\appendix
\section{Conventions}
\label{Conventions}
In this paper, we use the generators of $SL(2,\mathbb R)$
\begin{align} 
L_{-1}=\left(
  \begin{matrix}
   0 &\  0 \\
   1 &\  0 
  \end{matrix}
  \right),
  L_{0}=\frac{1}{2}\left(
  \begin{matrix}
   1 &\  0 \\
   0 &\  -1 
  \end{matrix}
  \right),
  L_{1}=\left(
  \begin{matrix}
   0 &\  1 \\
   0 &\  0 
  \end{matrix}
  \right).
\end{align}
The commutation relations are
\begin{align}
[L_{-1},L_0]=L_{-1},\quad [L_{-1},L_1]=-2L_0,\quad [L_{0},L_1]=L_1.
\end{align}
Its Cartan-Killing metric is
\begin{align}
\text{Tr}\left(L_iL_j\right)=&\left(
  \begin{matrix}
   0 &\  0 &\  1\\
   0 &\  \frac{1}{2}&\  0\\
   1 &\  0 &\  0\\
  \end{matrix}
  \right).
\end{align}
\section{Boundary term for $T\bar{T}$ deformation}
\label{variation TTbar}
In this appendix, we will derive the boundary term~\eqref{TT boundary term} for $T\bar T$ deformation. Firstly, we expect the variation of the total action behaves like the form  
\begin{align}
\delta S_{\rm{total}}=\frac{k}{4\pi}\int_{\partial M}d\tilde td\tilde\theta\text{Tr}\left[\left(\tilde{a}_{\tilde t}-\frac{1+\mu\mathcal {\bar L}_{\mu}}{1-\mu\mathcal {\bar L}_{\mu}}\tilde{a}_{\tilde\theta}\right)\delta \tilde a_{\tilde\theta}-\left(\bar {\tilde a}_{\tilde t}+\frac{1+\mu\mathcal { L}_{\mu}}{1-\mu\mathcal {L}_{\mu}}\bar {\tilde a}_{\tilde\theta}\right)\delta \bar {\tilde a}_{\tilde\theta}\right],
\end{align} 
which vanishes due to the mixed boundary condition. Therefore, the variation of boundary term can be identified as
\begin{align}
\label{boundary variation TT}
\delta \mathcal B=&-\frac{k}{4\pi}\int_{\partial M}d\tilde td\tilde\theta\left[\frac{1+\mu\mathcal {\bar L}_{\mu}}{1-\mu\mathcal {\bar L}_{\mu}}\text{Tr}\left(a_{\tilde\theta}\delta a_{\tilde\theta}\right)+\frac{1+\mu\mathcal {L}_{\mu}}{1-\mu\mathcal {L}_{\mu}}\text{Tr}\left(\bar a_{\tilde\theta}\delta \bar a_{\tilde\theta}\right)\right].
\end{align} 
According to~\eqref{a-tt} and~\eqref{abar-tt}, we can get the variation of $\tilde a,\tilde{\bar a}$ with respect to $\mathcal { L}_{\mu},\mathcal {\bar L}_{\mu}$
\begin{align}
\delta \tilde a_{\tilde\theta}=&\frac{1-\mu\mathcal {\bar L}_{\mu}}{(1-\mu^2\mathcal { L}_{\mu}\mathcal {\bar L}_{\mu})^2}(\mu^2\mathcal {\bar L}_{\mu}L_{-1}+L_1)\delta \mathcal{L}_{\mu}-\frac{\mu(1-\mu\mathcal {L}_{\mu})}{(1-\mu^2\mathcal { L}_{\mu}\mathcal {\bar L}_{\mu})^2}(L_{-1}+\mathcal{L}_{\mu}L_1)\delta \mathcal{\bar L}_{\mu},\\
\delta\tilde{\bar a}_{\tilde\theta}=&-\frac{\mu(1-\mu\mathcal {\bar L}_{\mu})}{(1-\mu^2\mathcal { L}_{\mu}\mathcal {\bar L}_{\mu})^2}(\mathcal {\bar L}_{\mu}L_{-1}+L_1)\delta \mathcal{L}_{\mu}+\frac{1-\mu\mathcal {L}_{\mu}}{(1-\mu^2\mathcal { L}_{\mu}\mathcal {\bar L}_{\mu})^2}(L_{-1}+\mu^2\mathcal {L}_{\mu}L_1)\delta \mathcal{\bar L}_{\mu}.
\end{align}
Besides, it is straightforward to obtain
\begin{align}
\text{Tr}\left(a_{\tilde\theta}\delta a_{\tilde\theta}\right)=&\frac{(1-\mu\mathcal {\bar L}_{\mu})^2(1+\mu^2\mathcal { L}_{\mu}\mathcal {\bar L}_{\mu})}{(1-\mu^2\mathcal { L}_{\mu}\mathcal {\bar L}_{\mu})^3}\delta \mathcal{L}_{\mu}-\frac{2\mu\mathcal{L}_{\mu}(1-\mu\mathcal {L}_{\mu})(1-\mu\mathcal {\bar L}_{\mu})}{(1-\mu^2\mathcal { L}_{\mu}\mathcal {\bar L}_{\mu})^2}\delta \mathcal{\bar L}_{\mu},\\
\text{Tr}\left(\bar a_{\tilde\theta}\delta \bar a_{\tilde\theta}\right)=&-\frac{2\mu\mathcal{\bar L}_{\mu}(1-\mu\mathcal {L}_{\mu})(1-\mu\mathcal {\bar L}_{\mu})}{(1-\mu^2\mathcal { L}_{\mu}\mathcal {\bar L}_{\mu})^3}\delta \mathcal{L}_{\mu}+\frac{(1-\mu\mathcal {L}_{\mu})^2(1+\mu^2\mathcal { L}_{\mu}\mathcal {\bar L}_{\mu})}{(1-\mu^2\mathcal { L}_{\mu}\mathcal {\bar L}_{\mu})^3}\delta \mathcal{\bar L}_{\mu}.
\end{align}
Substituting these relations into~\eqref{boundary variation TT}, it yields
\begin{align}
\delta \mathcal B=-\frac{\kappa}{2\pi}\int_{\partial M}d\tilde td\tilde\theta\left[\frac{(1-\mu\mathcal {\bar L}_{\mu})^2}{(1-\mu^2\mathcal { L}_{\mu}\mathcal {\bar L}_{\mu})^2}\delta \mathcal{L}_{\mu}+\frac{(1-\mu\mathcal {L}_{\mu})^2}{(1-\mu^2\mathcal { L}_{\mu}\mathcal {\bar L}_{\mu})^2}\delta \mathcal{\bar L}_{\mu}\right].
\end{align}
The right hand side of this equation is a total derivative. 
The expected primitive function of this boundary term variation could be 
\begin{align}
\label{boundary term 1}
\mathcal B=&-\frac{\kappa}{2\pi}\int_{\partial M}d\tilde td\tilde\theta\frac{\mathcal {L}_{\mu}+\mathcal {\bar L}_{\mu}-2\mu\mathcal {L}_{\mu}\mathcal {\bar L}_{\mu}}{1-\mu^2\mathcal { L}_{\mu}\mathcal {\bar L}_{\mu}}.
\end{align}
In addition, the boundary term could be written into another form
\begin{align}
\label{boundary term 2}
\mathcal B=&-\frac{\kappa}{4\pi}\int_{\partial M}d\tilde td\tilde\theta \left[\frac{1-\mu^2\mathcal { L}_{\mu}\mathcal {\bar L}_{\mu}}{1-\mu\mathcal {\bar L}_{\mu}}\text{Tr}\left(\tilde a_{\tilde\theta}^2\right)+\frac{1-\mu^2\mathcal { L}_{\mu}\mathcal {\bar L}_{\mu}}{1-\mu\mathcal {L}_{\mu}}\text{Tr}\left(\bar{\tilde a}_{\tilde\theta}^2\right)\right].
\end{align} 
As a consequence, the boundary term for $T\bar{T}$ deformation is just~\eqref{TT boundary term}. 
\section{Boundary term for $J\bar{T}$ deformation}
\label{variation JTbar}
In this appendix, we will derive the boundary term for $J\bar T$ deformation. According to the boundary condition~\eqref{a-jt},~\eqref{abar-jt} and~\eqref{phi-jt}, we can write down the expected variation of total action. We would like to consider the gravitational part and $U(1)$ gauge field part separately. For the gravitational action, its variation should take the following
\begin{align}
\delta S_{\rm{grav}}=&\frac{\kappa}{4\pi }\int_{\partial M}d\tilde td\tilde\theta\text{Tr}\left[\left(\tilde a_{\tilde{t}}-\tilde a_{\tilde \theta}\right)\delta \tilde{a}_{\tilde \theta}-\left(\bar{\tilde a}_{\tilde{t}}+\frac{1+\mu J}{1-\mu J}\bar{\tilde a}_{\tilde \theta}\right)\delta \bar {\tilde a}_{\tilde \theta}\right].
\end{align}
The variation of $U(1)$ gauge field action should be  
\begin{align}
\delta S_{\rm{U(1)}}=&\frac{\kappa'}{4\pi }\int_{\partial M}d\tilde td\tilde\theta\left(\tilde{\Phi}_{\tilde{t}}-\frac{4\pi}{\kappa'}\frac{J}{1-\mu J}+\frac{1+\mu J}{1-\mu J}\tilde \Phi_{\tilde \theta}\right)\delta\tilde\Phi_{\tilde \theta}.
\end{align}
Both of them vanish because of the boundary condition. Then, we can read off the variation of the boundary terms
\begin{align}
\label{JT boundary variation grav}
\delta \mathcal{B}_{\rm{grav}}=&-\frac{\kappa}{4\pi }\int_{\partial M}d\tilde td\tilde{\theta}\left[\text{Tr}(\tilde{a}_{\tilde \theta}\delta \tilde{a}_{\tilde \theta})+\frac{1+\mu J}{1-\mu J}\text{Tr}\left(\bar{\tilde a}_{\tilde \theta}\delta \bar {\tilde a}_{\tilde \theta}\right)\right],\\
\label{JT boundary variation U}
\delta \mathcal{B}_{\rm{U(1)}}=&-\frac{\kappa'}{4\pi }\int_{\partial M}d\tilde td\tilde\theta\left(\frac{4\pi}{\kappa'}\frac{J}{1-\mu J}-\frac{1+\mu J}{1-\mu J}\tilde \Phi_{\tilde \theta}\right)\delta\tilde\Phi_{\tilde \theta}.
\end{align}
By using~\eqref{a-jt},~\eqref{abar-jt} and~\eqref{phi-jt}, one can calculate 
\begin{align}
\text{Tr}\left(\tilde a_{\tilde \theta}\delta \tilde{a}_{\tilde \theta}\right)=&\delta \mathcal L,\\
\text{Tr}\left(\tilde a_{\tilde \theta}\delta \tilde{a}_{\tilde \theta}\right)=&(1-\mu J)^2\delta\bar{\mathcal {L}}_{\mu}-2\mu(1-\mu J)\bar{\mathcal {L}}_{\mu}\delta J,\\
\delta\tilde\Phi_{\tilde \theta}=&\left(\frac{2\pi}{\kappa'}-\mu\mathcal{F}\right)\delta J+(1-\mu J)\delta\mathcal{F}.
\end{align}
Plugging these relations into the boundary term and noting~\eqref{gauge field F}, we can write these boundary terms in terms of $\mathcal{L},J$ and $\mathcal{F}$  
\begin{align}
\delta \mathcal{B}_{\rm{grav}}=&-\int_{\partial M}d\tilde td\tilde\theta\frac{\kappa}{4\pi }\delta \mathcal L-\int_{\partial M}d\tilde td\tilde\theta(1-\mu^2 J^2)\left(\frac{1}{2\mu}-\frac{\kappa'}{4\pi}\mathcal F \right)\delta\mathcal F\nonumber\\
&+\int_{\partial M}d\tilde td\tilde\theta(1+\mu J)\left(\mathcal F-\frac{\mu\kappa'}{4\pi}\mathcal F^2 \right)\delta J,\\
\delta\mathcal{B}_{\rm{U(1)}}=&\int_{\partial M}d\tilde td\tilde\theta\left[\left(\frac{2\pi}{\kappa'}-\mu \mathcal F\right)^2\frac{\kappa'}{4\pi }J-\left(\frac{1}{2}-\frac{\kappa'\mu}{4\pi}\mathcal F\right)\mathcal F\right]\delta J\nonumber\\
&-\int_{\partial M}d\tilde td\tilde\theta\left[-\frac{1}{2}J(1-\mu J)+(1-\mu^2J^2)\frac{\kappa'}{4\pi }\mathcal F\right]\delta \mathcal F.
\end{align}
One can verify the variation of each boundary term is not a total derivative. However, combining the gravitational part and $U(1)$ gauge field part, we can get a total derivative. This might imply the boundary term coupling the gravity with $U(1)$ gauge field. The variation of total boundary term is 
\begin{align}
\label{variation of JT boundary term}
\delta \mathcal{B}=&\delta \mathcal{B}_{\rm{grav}}+\delta\mathcal{B}_{\rm{U(1)}}
=-\frac{\kappa}{4\pi }\int_{\partial M}d\tilde td\tilde\theta\left[\delta\mathcal L-\left(\frac{4\pi^2}{\kappa\kappa'}J+\frac{2\pi}{\kappa}\mathcal F\right)\delta J+\frac{2\pi}{\mu\kappa}(1-\mu J)\delta\mathcal F\right].
\end{align}
Integrate the above formula, we arrive at the expected boundary term
\begin{align}
\mathcal{B}=&-\frac{\kappa}{4\pi }\int_{\partial M}d\tilde td\tilde\theta\left[\mathcal L-\frac{2\pi^2}{\kappa\kappa'}J^2+\frac{2\pi}{\mu\kappa}(1-\mu J)\mathcal F\right].
\end{align}
\section{$J\bar T$ deformed WZW model}
\label{JT deformed WZW model}
In this appendix, we will derive one type of $J\bar T$ deformed chiral $SL(2,\mathbb R)$ WZW model from the $J\bar T$ flow equation, in which the $U(1)$ current is introduced by adding a left-moving chiral $U(1)$ WZW model action. We consider the action 
\begin{align} 
S_{\rm{total}}=&S_{\rm{LWZW}}^{SL(2,\mathbb R)}-S_{\rm{RWZW}}^{SL(2,\mathbb R)}+S_{\rm{LWZW}}^{U(1)}\nonumber\\
=&\int d^2x \mathscr L^{SL(2,\mathbb R)}_{\rm{LWZW}}+\Gamma[g]-\int d^2x\mathscr L^{SL(2,\mathbb R)}_{\rm{RWZW}}-\Gamma[\bar g]+\int d^2x \mathscr L^{U(1)}_{\rm{LWZW}}.
\end{align}
Here the Lagrangian for left-moving $SL(2,\mathbb R)$ WZW model is 
\begin{align}
\mathscr L^{SL(2,\mathbb R)}_{\rm{LWZW}}=&\frac{\kappa}{4\pi}\text{Tr}\left(\mathcal A_{\theta}\mathcal A_{t}-\mathcal A_{\theta}\mathcal A_{\theta}\right).
\end{align} 
In order to define the stress tensor, we put the right-moving $SL(2,\mathbb R)$ WZW model in a curved background whose metric is 
\begin{align}
g^{tt}=0,\quad g^{t\theta}=g^{\theta t}=\frac{1}{2},\quad g^{\theta\theta}=h.
\end{align}
Then the Lagrangian for right-moving $SL(2,\mathbb R)$ WZW model takes the form 
\begin{align}
\mathscr L^{SL(2,\mathbb R)}_{\rm{RWZW}}=&\frac{\kappa}{4\pi}\text{Tr}\left(\bar{\mathcal A}_{\theta}\bar {\mathcal A}_{t}+h\bar {\mathcal A}_{\theta}\bar {\mathcal A}_{\theta}\right).
\end{align}
In terms of the zweibeins, we can express $h$ as 
\begin{align}
h=\frac{e^{-}_t}{e^{-}_{\theta}}.
\end{align} 
Therefore, the Lagrangian for left-moving $SL(2,\mathbb R)$ WZW model can be written as  
\begin{align}
\mathscr L^{SL(2,\mathbb R)}_{\rm{RWZW}}=&\frac{\kappa}{4\pi}\text{Tr}\left(\bar {\mathcal A}_{\theta}\bar {\mathcal A}_{t}+\frac{e^{-}_{t}}{e^{-}_{\theta}}\bar{\mathcal A}_{\theta}\bar{\mathcal A}_{\theta}\right).
\end{align} 
This Lagrangian becomes chiral WZW action of left-moving copy if setting $h=-1$, and $h=1$ for the right-moving copy. We then couple $U(1)$ WZW model with gauge field $B$, such that the Lagrangian becomes
\begin{align}
\mathscr L^{U(1)}_{\rm{LWZW}}=\frac{\kappa'}{4\pi}\left[(\partial_{\theta}U\partial_{\theta}U-\partial_{\theta}U\partial_{t}U)+(B_{\theta}-B_t)(2\partial_{\theta}U+B_{\theta})\right].
\end{align}
Following the technique used for chiral Bosons~\cite{Sonnenschein:1988ug, Frishman:1987se,Ouyang:2020rpq}, we finally obtain the improved action
\begin{align} 
S_{\rm{imp}}
=&\frac{\kappa}{4\pi}\int d^2x\text{Tr}\left(\mathcal A_{\theta}\mathcal A_{t}-\mathcal A_{\theta}\mathcal A_{\theta}\right)+\Gamma[g]-\frac{\kappa}{4\pi}\int d^2x\text{Tr}\left(\bar {\mathcal A}_{\theta}\bar {\mathcal A}_{t}+\frac{e^{-}_{t}}{e^{-}_{\theta}}\bar{\mathcal A}_{\theta}\bar{\mathcal A}_{\theta}\right)-\Gamma[\bar g]\nonumber\\
&+\frac{\kappa'}{4\pi}\int d^2x[(\partial_{\theta}U\partial_{t}U-\partial_{\theta}U\partial_{\theta}U)-(B_{\theta}-B_t)(2\partial_{\theta}U+B_{\theta})].
\end{align}
Then the conserved stress tensor $\bar T^i_a$ and conserved current $J^i$ can be  defined by
\begin{align}
\bar T^{t}_{+}=&\frac{\partial\mathscr L}{\partial e^{+}_t},\quad \bar T^{\theta}_{+}=\frac{\partial\mathscr L}{\partial e^{+}_{\theta}}\\
J^t=&\frac{\partial \mathscr L}{\partial B_t},\quad J^{\theta}=\frac{\partial \mathscr L}{\partial B_{\theta}}
\end{align}
We identity this action as the original theory. 
\par
Therefore, the $J\bar T$-deformed Lagrangian $\mathscr L_{\mu}$ satisfy the flow equation
\begin{align}
\label{JT-flow}
\frac{\partial\mathscr L_{\mu}}{\partial \mu}=J^{t}\bar T^{\theta}_{+}-J^{\theta}\bar T^{t}_{+}=\frac{\partial \mathscr L_{\mu}}{\partial B_t}\frac{\partial\mathscr L_{\mu}}{\partial e^{+}_{\theta}}-\frac{\partial \mathscr L_{\mu}}{\partial B_{\theta}}\frac{\partial\mathscr L_{\mu}}{\partial e^{+}_{t}},
\end{align}
with the initial condition
\begin{align}
\mathscr L_{0}=\frac{\kappa}{4\pi}\text{Tr}\left(\mathcal A_{\theta}\mathcal A_{t}-\mathcal A_{\theta}\mathcal A_{\theta}\right)-\frac{\kappa}{4\pi}\text{Tr}\left(\bar {\mathcal A}_{\theta}\bar {\mathcal A}_{t}+\frac{e^{-}_{t}}{e^{-}_{\theta}}\bar{\mathcal A}_{\theta}\bar{\mathcal A}_{\theta}\right)\nonumber\\
+\frac{\kappa'}{4\pi}\left[(\partial_{\theta}U\partial_{\theta}U-\partial_{\theta}U\partial_{t}U)+(B_{\theta}-B_t)(2\partial_{\theta}U+B_{\theta})\right].
\end{align}
Solving the $J\bar{T}$ flow equation~\eqref{JT-flow}, and setting $e^{-}_t=e^{-}_{\theta}=1,B_{t}=B_{\theta}=0$, one can get the deformed Lagrangian 
\begin{align}
\mathscr L_{\mu}=&\frac{\kappa}{4\pi}\text{Tr}\left(\mathcal A_{\theta}\mathcal A_{t}-\mathcal A_{\theta}\mathcal A_{\theta}\right)-\frac{\kappa}{4\pi}\text{Tr}\left(\bar {\mathcal A}_{\theta}\bar {\mathcal A}_{t}+\bar{\mathcal A}_{\theta}\bar{\mathcal A}_{\theta}\right)+\frac{\kappa'}{4\pi}(\partial_{\theta}U\partial_{\theta}U-\partial_{\theta}U\partial_{t}U)\nonumber\\
&+\frac{2\pi}{\mu^2\kappa}\left(1-\frac{\mu\kappa'}{2\pi}\partial_{\theta}U-\sqrt{\left(1-\frac{\mu\kappa'}{2\pi}\partial_{\theta}U\right)^2+\frac{\mu^2\kappa\kappa'}{4\pi^2}\text{Tr}\left(\bar {\mathcal A}_{\theta}\bar {\mathcal A}_{\theta}\right)}\right).
\end{align}
Finally, the total action for $J\bar T$-deformed WZW model is 
\begin{align}
S_{J\bar{T}}=&\frac{\kappa}{4\pi}\int d^2x\text{Tr}\left(\mathcal A_{\theta}\mathcal A_{t}-\mathcal A_{\theta}\mathcal A_{\theta}\right)+\Gamma[g]\nonumber\\
&-\frac{\kappa}{4\pi}\int d^2x\text{Tr}\left(\bar {\mathcal A}_{\theta}\bar {\mathcal A}_{t}+\bar{\mathcal A}_{\theta}\bar{\mathcal A}_{\theta}\right)-\Gamma[\bar g]+\frac{\kappa'}{4\pi}\int d^2x(\partial_{\theta}U\partial_{t}U-\partial_{\theta}U\partial_{\theta}U)\nonumber\\
&+\frac{2\pi}{\mu^2\kappa}\int d^2x\left(1-\frac{\mu\kappa'}{2\pi}\partial_{\theta}U-\sqrt{\left(1-\frac{\mu\kappa'}{2\pi}\partial_{\theta}U\right)^2+\frac{\mu^2\kappa\kappa'}{4\pi^2}\text{Tr}\left(\bar {\mathcal A}_{\theta}\bar {\mathcal A}_{\theta}\right)}\right).
\end{align}

\providecommand{\href}[2]{#2}
\begingroup\raggedright

\endgroup

\end{document}